# Intelligence-Guided Adaptive Purification for DDoS-Resilient Quantum Networks: A CUDA-Q based Study

Santanu Ganguly[1,2,]
[1]*Kingston University London, Quantum AI Research Group, School of Computer Science and Mathematics*
*Penrhyn Road, Kingston upon Thames, KT1 2EE, United Kingdom*
[1]*e-mail address: : ganguly.santanu@kingston.ac.uk*
[2]*Endava plc, 125 Old Broad St, London EC2N 1AR, United Kingdom*
[2]*e-mail address: : santanu.gangoly@gmail.com*

Quantum-repeater networks require adaptive control policies that balance entanglement generation rate, end-to-end fidelity, purification overhead, and memory-induced latency. This tradeoff becomes more complex when the classical control plane is degraded by cyber anomalies or denial-of-service traffic. We develop a CUDA-Q/SeQUeNCe co-simulation workflow for studying adaptive entanglement purification in heterogeneous linear repeater chains. CUDA-Q noisy quantum kernels are used to estimate primitive entanglement purification and swapping behavior, while SeQUeNCe provides an event-layer model for stochastic link generation, waiting-time-dependent memory decay, purification failure, and end-to-end swapping. Under stationary conditions, we compare *no purification*, *local threshold purification*, *mean-field predictive purification*, *fixed purification*, and a *resource-penalized risk-aware predictive policy*. In an 8-node chain, the resource-penalized risk-aware controller increases above-target delivery probability relative to fixed purification while reducing latency and purification overhead. We then couple the quantum-network controller to anomaly scores derived from the CSE-CIC-IDS2018 benign-to-SSDP intrusion-detection trace. During the attack period, the attack-unaware controller maintains high raw delivery but its above-target entanglement delivery falls to $0.098 \pm 0.007$; the IDS-aware resource adaptive controller switches to more purification-heavy masks and increases above-target delivery to $0.344 \pm 0.011$, closely matching the oracle-aware value of ~0.335. These results demonstrate that cyber-state awareness can improve the useful quantum-network outcome by trading raw throughput for fidelity-qualified entanglement delivery.

**Keywords:** quantum machine learning; quantum networks; cybersecurity; anomaly detection; monte carlo; CUDA-Q.

## I. INTRODUCTION

Quantum networks aim to distribute entanglement between remote nodes for quantum key distribution, distributed sensing, blind or delegated computation, and ultimately networked quantum processors [1,2]. Because unknown quantum states cannot be copied, long-distance quantum communication cannot rely on simple optical amplification. Quantum repeaters instead use elementary entanglement generation, quantum memories, entanglement purification, and entanglement swapping to extend high-fidelity entanglement across lossy links [3–5]. In realistic networks, these operations are probabilistic and resource-constrained: elementary links may fail to generate, purification consumes pairs and may fail, memories decohere during synchronization, and swapping compounds local gate and measurement errors. More broadly, high-fidelity distribution over lossy and probabilistic links depends on these same primitives [6-10].

A repeater-chain controller must balance raw delivery probability, final fidelity, purification overhead, memory decay, and latency [2-7]. Maximizing raw delivery alone can produce many low-fidelity pairs, whereas purifying every elementary link can increase failure probability and delay. Adaptive purification is therefore naturally a rate-fidelity-resource control problem [15, 16, 46-48]. Quantum machine learning has also become a systematic direction for quantum-network control and anomaly detection, including quantum autoencoders and variational quantum circuits [12–17, 76, 77], with recent extensions to mixed-state, time-series, and cybersecurity anomaly-detection settings [19–27], and space based data centers (SBDC) 78].

This paper addresses a central control question in such networks: when should a repeater purify an elementary link, and when should it preserve delivery rate by avoiding additional purification overhead? A local threshold rule purifies links whose raw fidelity falls below a fixed trigger, but does not capture how local fidelity changes propagate through an end-to-end chain. Similarly, fixed purification can improve delivered fidelity but may overuse resources and increase purification failure. We therefore study predictive policies that optimize end-to-end useful delivery, defined as the probability of delivering an end-to-end entangled pair whose final fidelity exceeds a target threshold [11, 13, 28-31].

We study this problem using a CUDA-Q/SeQUeNCe co-simulation workflow. CUDA-Q noisy quantum kernels estimate primitive purification and swapping behavior, providing circuit-level estimates of the quantum operations used by the repeater protocol. SeQUeNCe-style event-layer simulation models stochastic link generation, retry limits, memory decay, purification failure, latency, memory waiting times, swapping, and end-to-end entanglement delivery [11, 28-31]. This separation allows noisy quantum primitives to be evaluated at the circuit level while network-level policies are tested over many Monte Carlo trials, aligning with the broader use of discrete-event and event-layer simulators such as SeQUeNCe and NetSquid for quantum-network protocol evaluation [11–14]. Under stationary conditions, we compare no purification, threshold-adaptive purification, mean-field predictive purification, fixed purification, and resource-penalized risk-aware purification in three-link and eight-node repeater chains [5-7].

Finally, we link cybersecurity and quantum-network control by studying how adaptive purification should respond when the classical control environment becomes anomalous. Quantum networks are cyber-physical systems because the quantum data plane depends on a classical control plane for heralding, synchronization, purification decisions, routing, and feedforward [1–6, 45–48, 60, 63, 65, 80]. Cyber anomalies such as denial-of-service and intrusion events in the classical layer can therefore affect quantum-network operation and performance by increasing latency, degrading control reliability, degrading effective link generation, or reducing final entanglement fidelity [35-39]. To model this, we construct a benign-to-SSDP anomaly trace from the CSE-CIC-IDS2018 dataset [32-34], map the cyber signal to time-varying quantum-link degradation [83], and compare clean, attack-unaware, IDS-aware (Intrusion Detection System-aware), and oracle-aware quantum-network controllers. The result is a unified workflow in which hybrid quantum anomaly analysis provides interpretable cyber-state information, and IDS-derived anomaly scores become actionable inputs for adaptive purification in quantum-repeater networks.

The results support two central conclusions. First, under stationary network conditions, resource-penalized risk-aware purification can outperform fixed purification in the 8-node chain by reducing over-purification and avoiding excess purification failures. Second, under SSDP-driven degradation [41-44], IDS-aware control improves the probability of useful above-target entanglement delivery by switching to more purification-heavy masks during the attack period.

**Structure of the paper**: Section II describes the motivation and contribution; Section III the methodology; Section IV presents and analyses the results followed by the conclusion in Section V. Appendix A presents supporting results followed by Appendix B presenting theory.

# II. MOTIVATION AND CONTRIBUTION

The *motivation* for this work is three-fold:

*First*, adaptive purification in quantum-repeater chains requires optimizing useful entanglement delivery rather than raw delivery alone. Raw delivery counts every end-to-end pair that arrives, whereas useful delivery counts only those pairs whose final fidelity exceeds a target threshold. Under stationary conditions, a fixed purification rule may overuse resources, while a no-purification rule may deliver too many below-target pairs. Under degraded conditions, this distinction becomes even sharper.

*Second*, predictive purification should be evaluated under stochastic event-layer conditions. Analytical Werner-state recursions are valuable for intuition, but event-layer behavior depends on waiting times, finite retry budgets, memory decay, link heterogeneity, and correlated failure modes. These effects become especially important in longer chains.

*Third*, quantum networks require a trusted and responsive classical control layer [1–7, 45–48, 79]. If the control environment is under attack, a quantum-network policy should not treat its link parameters as stationary. IDS-derived anomaly scores can serve as cyber-state inputs, allowing the controller to adapt purification decisions when attack-driven degradation threatens final entanglement fidelity [32–40].

The *central operational tradeoff* is that IDS-aware control may deliberately sacrifice raw delivery probability in order to increase above-target delivery probability. In other words, the goal is not simply to deliver more entangled pairs, but to deliver more fidelity-qualified entangled pairs.

The quantum-network portion builds on the standard repeater toolkit of entanglement generation, purification, memory storage, and entanglement swapping [1–7]. Discrete-event simulators such as SeQUeNCe and NetSquid make it possible to evaluate such protocols beyond purely analytical models by accounting for latency, classical control, memory decay, and stochastic event trajectories [31–34]. CUDA-Q provides the noisy quantum-kernel layer used here for primitive purification and swapping estimates [35, 36].

The cyber-aware control portion is the bridge between the two. Existing IDS datasets and anomaly-detection algorithms are typically evaluated by classification or ranking metrics [32–44]. Existing quantum-network simulations are typically evaluated under stationary or externally specified link conditions [11–14, 29]. Here, IDS-derived anomaly severity becomes an input to the quantum-network controller, allowing cyber-state awareness to change the purification policy itself.

**Contributions**: *First*, we develop a CUDA-Q/SeQUeNCe co-simulation workflow for adaptive purification in quantum-repeater chains. CUDA-Q noisy kernels estimate purification and swapping primitives, while the event-layer simulator evaluates end-to-end delivery, latency, memory decay, purification failure, and final fidelity over many Monte Carlo trials.

*Second*, we compare no purification, threshold-adaptive purification, mean-field predictive purification, fixed purification, and risk-aware predictive purification under stationary network conditions.

*Third*, we introduce a resource-penalized risk-aware purification objective for repeater-chain control. The objective combines a lower-confidence-bound estimate of above-target delivery, empirical useful delivery, delivered-fidelity margin, purification cost, and latency cost.

*Fourth*, we show that in an eight-node stationary chain, the resource-aware controller improves above-target delivery relative to fixed purification while reducing latency and purification overhead.

*Fifth*, we couple IDS DDoS data and the quantum-network components through an anomaly-aware control experiment. A benign-to-SSDP trace from CSE-CIC-IDS2018 is converted into an IDS severity signal and an oracle attack-rate signal. The physical network degradation follows the oracle attack rate, while the IDS-aware controller plans using the learned severity score. During the SSDP attack period, IDS-aware control increases the above-target entanglement-delivery probability from $0.098 \pm 0.007$ for the attack-unaware controller to $0.344 \pm 0.011$, comparable to the oracle-aware value of $0.335 \pm 0.011$, with uncertainties denoting 95% confidence half-widths over Monte Carlo trials.

The combined study provides a cyber-aware quantum-network workflow and establishes a link between quantum-enhanced anomaly analysis and quantum-network control rather than treating them as unrelated demonstrations.

# III. METHODS

## A. Part 1: Adaptive purification under stationary quantum-network conditions

**CUDA-Q primitive validation:** We first validated the elementary quantum operations used by the repeater-chain model using CUDA-Q noisy quantum kernels [28, 29]. The goal of this stage was not to simulate the full repeater network as one large quantum circuit. Instead, CUDA-Q was used to estimate and validate the primitive operations that enter the event-layer network simulation: (1) Bell-pair generation; (2) entanglement purification, and (3) entanglement swapping.

The Bell-pair validation circuit prepares the Bell state

$$\left|\Phi^{+}\right\rangle = \frac{|00\rangle + |11\rangle}{\sqrt{2}} \tag{1}$$

This was implemented with a Hadamard gate followed by a controlled-NOT gate. Sampling the circuit verifies that the CUDA-Q target and measurement convention produce approximately equal populations in the 00 and 11 outcomes, with negligible odd-parity outcomes in the ideal case.

The entanglement-swapping validation circuit prepares two elementary Bell pairs, applies a Bell-basis measurement on the two repeater qubits, and estimates the Bell-state fidelity of the two endpoint qubits. Because different simulation backends can expose measurement bitstrings in different orders, the mapping between the measured Bell-basis outcomes and the corresponding Pauli-frame correction was calibrated before running the network experiments. The calibrated mapping was then held fixed.

For purification, a BBPSSW-style two-pair primitive was used [4-10]. Two imperfect Bell pairs are treated as Werner-like input states with fidelities $F_1$ and $F_2$. A bilateral CNOT operation is applied, followed by measurement and postselection. The purification primitive returns two quantities. The *first* is the purification success probability:

$$p_{\text{pur}}\left(F_1, F_2\right).$$

This is the probability that the purification attempt succeeds. The *second* is the conditional output fidelity:

$$F_{\text{pur}}\left(F_1, F_2\right).$$

This is the fidelity of the purified pair, conditioned on the purification attempt being successful. The swapping primitive similarly returns an estimated output fidelity:

$$F_{\text{swap}}\left(F_1, F_2\right).$$

This is the end-to-end fidelity obtained by swapping two input links with fidelities $F_1$ and $F_2$.

The CUDA-Q primitive checks verified three things. First, ideal Bell-pair preparation and ideal entanglement swapping produced near-unit endpoint fidelity. Second, ideal purification produced near-unit success probability and near-unit conditional output fidelity. Third, noisy input fidelities reduced output fidelity in the expected direction. These checks establish that the primitive estimators are internally consistent before they are embedded into the event-layer Monte Carlo network simulation.

To reduce runtime, primitive estimates were cached. When the event-layer simulator requested a purification or swapping estimate for a given pair of input fidelities, the fidelities were rounded to a fixed grid. If the same rounded fidelity pair had already been evaluated, the cached CUDA-Q estimate was reused. This does not change the repeater protocol; it only avoids repeatedly sampling the same noisy quantum circuit.

In this section, we follow the layout depicted in FIG. 1, where, for example, a purification mask such as $m = 101$ in the 3-node baseline network of FIG 1(a), would mean: *purify link 1(L1)*, *do not purify link 2(L2)*, *purify link 3(L3)*;

similarly, a mask of $m = 1111011$ in the 8-node network of FIG. 1(b) would indicate *L1 purified, L2 purified, L3 purified, L4 purified, L5 not purified, L6 purified,* and *L7 purified.*

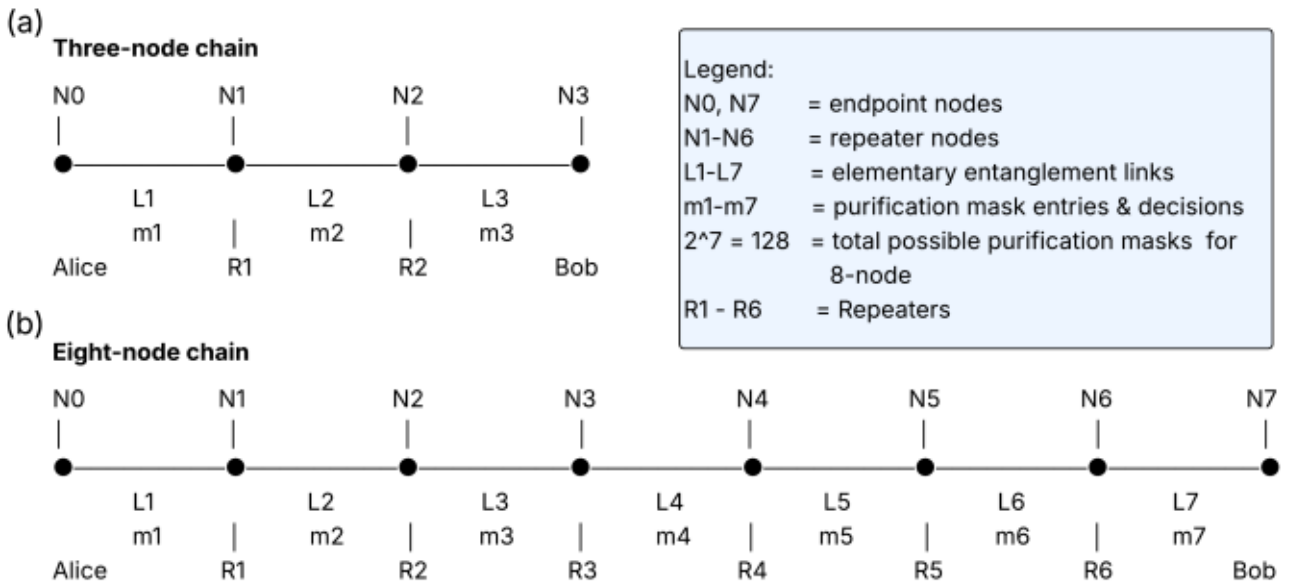

Step 1: Generate elementary Bell pairs on L1,...,L7
Step 2: Apply purification where mi = 1
Step 3: Perform entanglement swapping at repeater nodes N1,...,N6
Step 4: Deliver final end-to-end pair between N0 and N7

FIG. 1. (Colour online) *Linear repeater-chain topologies used in the stationary adaptive-purification experiments*. (a) The three-link baseline contains four nodes, two repeater nodes, and three elementary links. (b) The eight-node chain contains seven elementary links and six repeater nodes. Each elementary link $L_i$ is assigned a purification-mask entry $m_i \in \{0,1\}$, where $m_i = 1$ denotes purification of link $i$. For the eight-node chain, the controller evaluates $2^7 = 128$ candidate purification masks.

**SeQUeNCe event-layer simulation:** The repeater network was modeled as a linear chain of $N$ nodes. The number of elementary links is $L = N - 1$.

Each elementary link $i$ is described by four parameters:

$$\left(p_i, F_i, \tau_i, T_{2,i}\right).$$

Here, $p_i$ is the probability of successful elementary entanglement generation per attempt, $F_i$ is the raw Bell-pair fidelity, $\tau_i$ is the duration of one generation attempt, and $T_{2,i}$ is the memory coherence time. The simulation follows a discrete-event structure similar to SeQUeNCe and NetSquid [11, 13, 80, 81]. A request begins by attempting to generate all elementary links. Each link is attempted until it succeeds or until a maximum retry budget is reached. If any elementary link fails to generate within the allowed number of attempts, the request terminates as a link-generation failure.

Because different links may be generated at different times, successfully generated quantum memories may have to wait until the rest of the chain is ready. During this waiting time, memory decay reduces the stored fidelity. If a link with initial fidelity $F(0)$ waits for time $t$, the decayed fidelity is modelled as

$$F(t) = \frac{1}{4} + \left(F(0) - \frac{1}{4}\right)\exp\left(-\frac{t}{T_2}\right). \qquad (2)$$

This formula means that, as waiting time increases, the Bell-pair fidelity relaxes toward $1/4$, the maximally mixed Bell-state value. A purification policy is represented by a binary mask:

$$m = \left(m_1, m_2, \ldots, m_L\right).$$

Each element satisfies: $m_i \in \{0,1\}$.

The interpretation is:

$$m_i = \begin{cases} 1, & \text{purify elementary link } i, \\ 0, & \text{do not purify elementary link } i. \end{cases}$$

The total number of planned purifications is

$$|m| = \sum_{i=1}^{L} m_i.$$

If purification is selected for link $i$, the simulator samples a purification-success event using the CUDA-Q-estimated probability

$$p_{\text{pur}}\left(F_1, F_2\right).$$

If purification succeeds, the link fidelity is updated to

$$F_{\text{pur}}\left(F_1, F_2\right).$$

If purification fails, the request terminates with status `purification_failed`. After all selected purifications are completed, the elementary links are connected by entanglement swapping [5]. Each swapping operation is sampled with a fixed success probability, and the running end-to-end fidelity is updated using the CUDA-Q-estimated swapping primitive:

$$F_{\text{swap}}\left(F_1, F_2\right).$$

If any swapping operation fails, the request terminates with status swapping_failed. If all swapping operations succeed, the request is delivered with final end-to-end fidelity

$$F_{\text{end}}.$$

The raw delivery probability is

$$P_{\text{del}} = \Pr\left(D = 1\right),$$

where $D = 1$ means that an end-to-end pair was successfully delivered.

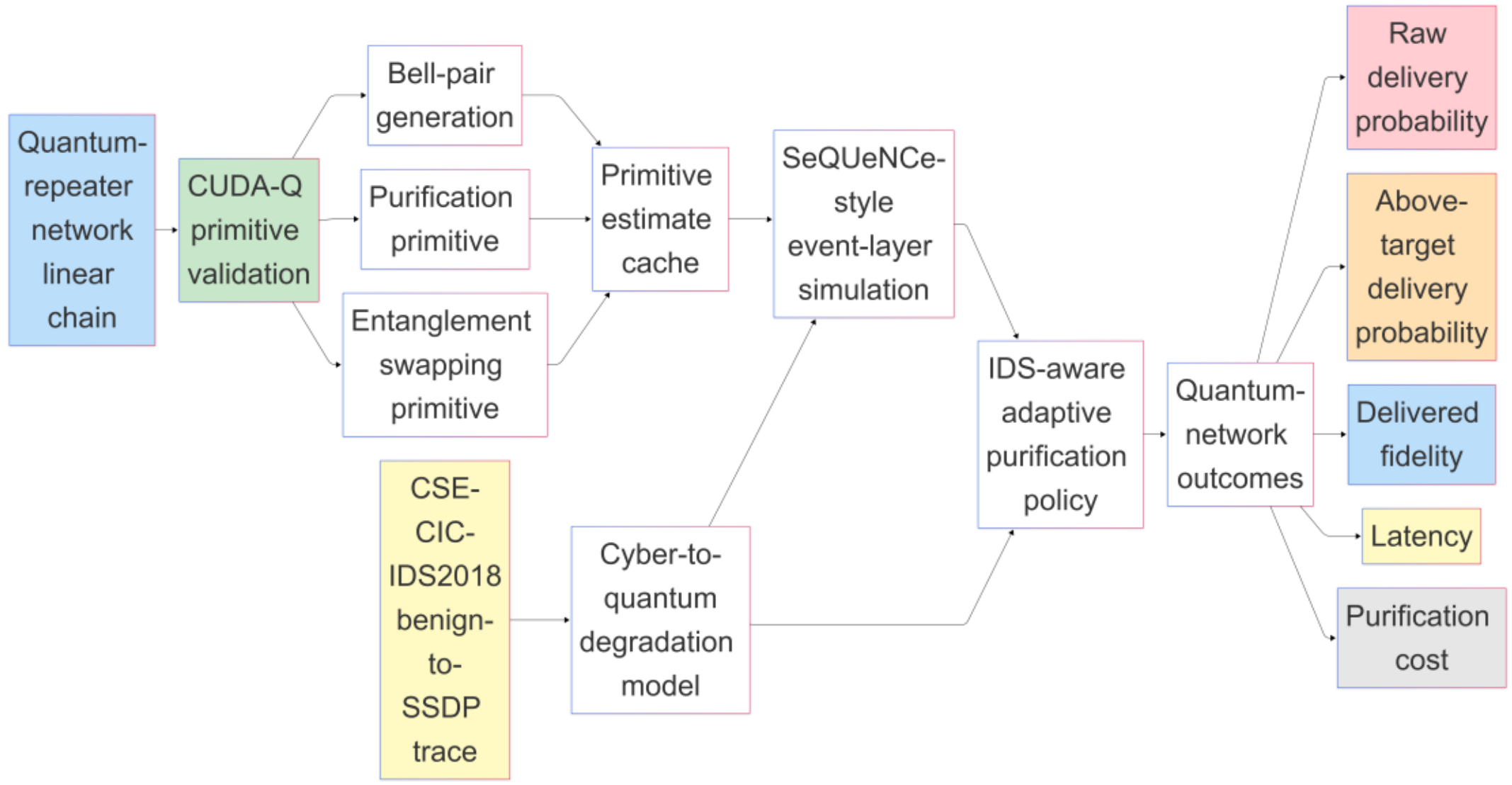

FIG. 2. (Colour online) *Event layer simulation workflow*. (a) CUDA-Q noisy kernels validate the primitive Bell-pair, purification, and entanglement-swapping operations and provide cached estimates of purification success probability, conditional purification fidelity, and swapping output fidelity. (b) These primitive estimates are embedded into a SeQUeNCe-style event-layer simulation of a linear repeater chain with stochastic link generation, memory waiting-time decay, purification failure, and entanglement swapping. (c) A three-link chain is used as a controlled baseline for comparing no purification, threshold-adaptive purification, predictive adaptive purification, and fixed purification. (d) The model is then extended to an eight-node chain, where all $2^7 = 128$ purification masks are evaluated by an inner Monte Carlo planner and a resource-aware predictive policy selects the mask that balances above-target delivery, delivered fidelity, purification cost, and latency. Full model pipeline is shown in Fig. 11 in Appendix A.

The main useful-delivery metric is the above-target delivery probability:

$$U = \Pr\left(D = 1 \text{ and } F_{\text{end}} \geq F_\star\right), \tag{3}$$

where $F_\star$ is the target fidelity threshold. Thus, raw delivery counts every delivered pair, while above-target delivery counts only delivered pairs whose final fidelity exceeds the required target.

For Bernoulli metrics such as raw delivery probability and above-target delivery probability, the 95% confidence half-width was computed as

$$\Delta_{95} = 1.96\sqrt{\frac{\hat{p}\left(1-\hat{p}\right)}{n}}.$$

Here, $\hat{p}$ is the empirical probability from Monte Carlo simulation, and $n$ is the number of trials.

For continuous delivered quantities, such as delivered fidelity and delivered latency, the 95% confidence half-width was computed using the standard error of the delivered samples.

The high-level model pipeline is shown in FIG. 2., with the full pipeline share in FIG. 11 of the Appendix A.

**Three-link baseline:** We first evaluated the adaptive-purification policies on a three-link repeater chain. This smaller chain was used as a controlled baseline because the number of possible purification masks is small and the rate-fidelity tradeoff is easy to inspect.

The three-link chain has four nodes and three elementary links. Since $L = 3$, there are $2^3 = 8$ possible purification masks. In total, *four policies* were compared:

*a. No purification***:** No elementary link is purified. This policy tends to maximize raw delivery probability because it avoids purification failures and additional latency. However, it may deliver end-to-end pairs whose final fidelity is below the target.

*b. Threshold-adaptive purification***:** An elementary link is purified if its raw or predicted link fidelity falls below a fixed purification trigger. This is a local rule. It is simple, but it does not directly optimize the final end-to-end fidelity.

*c. Predictive adaptive purification***:** Candidate purification masks are evaluated using predicted end-to-end fidelity and delivery probability. The controller selects a mask that is predicted to satisfy the target-fidelity requirement with lower resource cost. If no mask can satisfy the target, the best available mask is selected.

*d. Fixed purification***:** Every elementary link is purified. This policy increases local link fidelity but can reduce delivery probability because every purification attempt introduces an additional failure mode and additional latency.

The three-link baseline was used to identify operating regimes. At low target fidelities, no purification can be sufficient. At intermediate target fidelities, predictive adaptive purification can improve useful above-target delivery without the full cost of purifying every link. At high target fidelities, fixed purification can produce higher delivered fidelity, but at the cost of lower delivery probability and increased latency.

For each policy, we report: (i) $P_{\text{del}}$, the raw delivery probability; (ii) $U$, the above-target delivery probability; (iii) $\bar{F}_{\text{del}}$, the mean delivered fidelity; (iv) $\bar{L}_{\text{del}}$, the mean delivered latency, and (v) $\bar{N}_{\text{pur}}$, the mean number of purification operations.

**Eight-node risk-aware purification:** We then extended the analysis to an eight-node linear chain. Consider a linear quantum-repeater chain with $N$ nodes. The number of elementary links is

$$L = N - 1.$$

A purification policy is represented by a binary mask

$$m = (m_1, m_2, \ldots, m_L),$$

where each entry satisfies $m_i \in \{0,1\}$. The interpretation is,

$$m_i = \begin{cases} 1, & \text{purify elementary link } i, \\ 0, & \text{do not purify elementary link } i. \end{cases}$$

The total number of planned purification operations is

$$|m| = \sum_{i=1}^{L} m_i.$$

For an 8-node linear chain,

$$N = 8,\ L = 7,$$

so there are $2^7 = 128$ possible purification masks. This is small enough for exhaustive mask evaluation but large enough to show nontrivial resource allocation across heterogeneous links.

For the eight-node chain, we introduced a resource-penalized risk-aware predictive policy. For each candidate purification mask $m$, an inner Monte Carlo planner estimates:

- above-target delivery probability,
- raw delivery probability,
- delivered fidelity,
- delivered latency, and
- purification count.

*a. Delivery and above-target utility*: For a given purification mask $m$, let $D_m \in \{0,1\}$ denote whether the end-to-end entangled pair is successfully delivered.

Let $F_m$ denote the final end-to-end fidelity of the delivered pair and $F_\star$ be the target fidelity threshold. Then, the raw delivery probability is

$$P_{\text{del}}(m) = \Pr(D_m = 1).$$

The useful or above-target delivery probability is

$$U(m) = \Pr(D_m = 1 \text{ and } F_m \geq F_\star). \tag{4}$$

This is the main metric in the paper, because it counts only end-to-end pairs that are both delivered and above the required fidelity threshold.

The mean delivered fidelity is

$$\bar{F}_{\text{del}}(m) = \mathbb{E}[F_m \mid D_m = 1].$$

The mean delivered latency is

$$\bar{L}_{\text{del}}(m) = \mathbb{E}[L_m \mid D_m = 1],$$

where $L_m$ is the total latency of the delivered request.

*b. Empirical Monte Carlo estimators*: Suppose we run $n$ independent Monte Carlo trials for a fixed mask $m$. The Bernoulli random variable is defined as,

$$Y_j(m) = 1\left\{D_m^{(j)} = 1 \text{ and } F_m^{(j)} \geq F_\star\right\},$$

where $j$ indexes the Monte Carlo trial. Then the empirical above-target delivery probability estimated during planning is given by,

$$\hat{U}_n(m) = \frac{1}{n}\sum_{j=1}^{n} Y_j(m).$$

Because $Y_j(m)$ is Bernoulli, a standard 95% confidence half-width is

$$\Delta_{95}(m) = 1.96\sqrt{\frac{\hat{U}_n(m)\left(1 - \hat{U}_n(m)\right)}{n}}.$$

The lower-confidence-bound (LCB) term is

$$\hat{U}_{\text{LCB}}(m) = \hat{U}(m) - 1.96\sqrt{\frac{\hat{U}(m)\left(1 - \hat{U}(m)\right)}{n_{\text{plan}}}}. \tag{5}$$

This rewards masks that perform well while accounting for Monte Carlo uncertainty. Thus, the reported confidence interval is

$$\hat{U}_n(m) \pm \Delta_{95}(m).$$

The same form is used for raw delivery probability.

*c. Risk-aware resource-penalized objective*: The risk-aware controller evaluates all purification masks and chooses the one with the best score.

If $\hat{U}_{\mathrm{LCB}}(m)$ is the lower-confidence-bound estimate of above-target delivery:

$$\hat{U}_{\mathrm{LCB}}(m) = \hat{U}_n(m) - \Delta_{95}(m). \tag{6}$$

The resource-penalized planning score is

$$J(m) = \hat{U}_{\mathrm{LCB}}(m) + \alpha \hat{U}_n(m) + \gamma G(m) - \lambda_p |m| - \lambda_\ell \hat{L}_{\mathrm{del}}(m).$$

Here, $\alpha \geq 0$ weights the empirical above-target delivery probability; $\gamma \geq 0$ weights the delivered-fidelity margin, $\lambda_p \geq 0$ penalizes the number of purification operations, and $\lambda_\ell \geq 0$ penalizes delivered latency. The delivered fidelity-margin term is

$$G(m) = \mathrm{clip}\left(\hat{F}_{\mathrm{del}}(m) - F_\star, -g_{\max}, g_{\max}\right). \tag{7}$$

This term rewards masks whose delivered fidelity exceeds the target $F_\star$. The clipping prevents extremely large positive or negative fidelity margins from dominating the score. The term $\lambda_p |m|$ penalizes purification cost. The term $\lambda_\ell \hat{L}_{\mathrm{del}}(m)$ penalizes latency. The selected mask is

$$\hat{m} = \arg \max_{m \in \{0,1\}^7} J(m).$$

This objective was designed to avoid two extremes. A policy that focuses only on raw delivery tends to choose too little purification and may deliver many below-target pairs. A policy that purifies every link may improve fidelity but can lose too many requests to purification failure and increased latency. The risk-aware objective instead optimizes useful above-target delivery while explicitly accounting for purification and latency costs.

The eight-node stationary experiments compared:

1. no purification,
2. threshold-adaptive purification,
3. mean-field predictive purification,
4. risk-aware predictive purification, and
5. fixed purification.

The main question was whether the risk-aware policy could improve useful delivery relative to fixed purification by avoiding over-purification while still preserving sufficient fidelity margin.

For each policy, we report: (i) $P_{\mathrm{del}}$, the raw delivery probability; (ii) $U$, the above-target delivery probability; (iii) $\bar{F}_{\mathrm{del}}$, the mean delivered fidelity; (iv) $\bar{L}_{\mathrm{del}}$, the mean delivered latency, (v) $\bar{N}_{\mathrm{attempts}}$, mean number of generation attempts and (vi) $\bar{N}_{\mathrm{pur}}$, the mean number of purification operations.

Failure-stage statistics were also recorded, including purification failures and swapping failures. These statistics help explain whether a policy fails because it purifies too aggressively, swaps too unreliably, or delivers below-target pairs.

Target-fidelity sweeps and purification-trigger sweeps were then performed to test how the policy ordering changes across operating regimes.

Further theoretical details are presented in Appendix B.

## B. Part 2: IDS-aware quantum-network control

**CSE-CIC-IDS2018 benign-to-SSDP trace:** To couple cyber-state information to the quantum-network controller, we constructed a benign-to-SSDP anomaly trace from the CSE-CIC-IDS2018 intrusion-detection dataset [4,5]. The purpose of this trace was not to train the final quantum-network policy directly, but to provide a time-dependent cyber signal that could be used to perturb the quantum-repeater model and to inform adaptive purification decisions.

The CSE-CIC-IDS2018 dataset contains labeled flow-level traffic generated using CICFlowMeter-style features. We used a subset containing benign traffic and SSDP attack traffic. Each flow is represented by a vector of traffic features, such as packet counts, flow duration, packet-length statistics, inter-arrival-time features, and bidirectional traffic summaries. The ground-truth label identifies whether a flow is benign or belongs to the SSDP attack class.

The flow table was divided into consecutive time bins. In each bin, we computed two cyber quantities. *First*, we computed an IDS anomaly score, $a_{\mathrm{IDS}}(t)$, where $t$ is the index of the time bin. This score was obtained by fitting an anomaly-detection model to benign traffic and then applying it to the benign-to-SSDP trace. The score was normalized so that

$$0 \leq a_{\mathrm{IDS}}(t) \leq 1.$$

Values near zero correspond to benign-like traffic, while values near one correspond to strongly anomalous traffic.

Second, we computed an oracle attack-rate signal, $a_{\mathrm{oracle}}(t)$, using the ground-truth labels. For a given time bin $t$,

$$a_{\mathrm{oracle}}(t) = \frac{\text{number of attack flows in bin } t}{\text{total number of flows in bin } t}.$$

Thus, $a_{\mathrm{oracle}}(t) = 0$ means the bin is entirely benign, and $a_{\mathrm{oracle}}(t) = 1$ means the bin is entirely attack traffic.

The IDS-aware controller only uses $a_{\mathrm{IDS}}(t)$. The oracle signal $a_{\mathrm{oracle}}(t)$. is used for two purposes: to define the physical degradation applied to the network in the controlled experiment, and to provide an upper-reference controller that knows the true attack rate. This produces a time-resolved cyber trace of the form

$$\{a_{\text{IDS}}(t), a_{\text{oracle}}(t)\}_{t=1}^{T}.$$

The trace begins in a benign regime, transitions into an SSDP attack regime, and remains in a high-severity attack regime after the transition. This structure allows us to test whether a quantum-network controller can adapt its purification behavior when the classical cyber environment changes.

**Cyber-to-quantum degradation model:** The cyber-to-quantum degradation model maps the anomaly trace into time-dependent changes in selected quantum-link parameters. The goal is to represent a situation in which an attack on the classical control environment indirectly degrades the performance of the quantum-repeater network.

The baseline quantum-repeater chain has elementary links indexed by

$$i = 1, 2, \ldots, L.$$

Each link $i$ has four stationary parameters:

$$p_i^{(0)}, F_i^{(0)}, \tau_i^{(0)}, T_{2,i}^{(0)}.$$

Here:

- $p_i^{(0)}$ is the baseline link-generation probability,
- $F_i^{(0)}$ is the baseline raw Bell-pair fidelity,
- $\tau_i^{(0)}$ is the baseline attempt time, and
- $T_{2,i}^{(0)}$ is the baseline memory coherence time.

During an attack, a subset of links is treated as affected by the cyber anomaly. For an affected link, the degradation level is controlled by a scalar anomaly severity

$$a(t) \in [0,1].$$

In the physical degradation model, we use the oracle attack-rate signal,

$$a(t) = a_{\text{oracle}}(t),$$

so that the simulated physical degradation follows the actual labeled attack intensity.

For an affected link, the degraded parameters are

$$p_i(t) = p_i^{(0)}\left(1 - \eta_p a(t)\right),\ F_i(t) = F_i^{(0)} - \eta_F a(t),$$

$$\tau_i(t) = \tau_i^{(0)}\left(1 + \eta_\tau a(t)\right),$$

$$T_{2,i}(t) = T_{2,i}^{(0)}\left(1 - \eta_T a(t)\right).$$

The constants $\eta_p, \eta_F, \eta_\tau, \eta_T$ control the strength of degradation. The interpretation is as follows: A larger anomaly score reduces the probability of successful elementary entanglement generation:

$$p_i(t) < p_i^{(0)}.$$

It also reduces the raw link fidelity:

$$F_i(t) < F_i^{(0)}.$$

It increases the effective attempt time:

$$\tau_i(t) > \tau_i^{(0)}.$$

Finally, it reduces the effective memory coherence time:

$$T_{2,i}(t) < T_{2,i}^{(0)}.$$

This degradation model is intentionally phenomenological. It does not claim that SSDP packets directly modify quantum hardware. Instead, it models the operational effect of a degraded classical control environment on a quantum-repeater network. In such an environment, control-plane congestion or attack-driven disruption may reduce synchronization reliability, increase effective latency, and lower the quality of delivered entanglement.

The IDS-aware controller does not know the oracle degradation signal. Hence, it plans using the learned IDS severity signal, $a_{\text{IDS}}(t)$, while the physical network is degraded according to $a_{\text{oracle}}(t)$. This separation allows us to compare learned cyber awareness against an oracle controller.

**Attack-unaware, IDS-aware, and oracle-aware control:** For each time bin, we evaluate four scenarios.

*a. Clean counterfactual*: The clean counterfactual uses the baseline quantum links without cyber degradation. That is,

$$a(t) = 0.$$

The controller selects a purification mask using the clean network parameters. This scenario represents the performance that would be expected if the SSDP attack did not affect the quantum-network environment.

*b. Attack-unaware control***:** In the attack-unaware scenario, the physical quantum network is degraded according to the oracle attack rate, $a(t) = a_{\text{oracle}}(t)$, but the controller does not adapt to the attack. It continues to use the purification mask selected under clean conditions.

This scenario tests what happens when the network experiences attack-driven degradation but the purification controller assumes stationary conditions.

The expected behavior is that raw delivery may remain relatively high, because the controller does not increase purification overhead. However, the final delivered fidelity may drop, causing many delivered pairs to fall below the target fidelity threshold.

*c. IDS-aware control*: In the IDS-aware scenario, the physical quantum network is still degraded according to $a_{\text{oracle}}(t)$, but the controller plans using the learned IDS severity score, $a_{\text{IDS}}(t)$. The IDS-aware controller first constructs an estimated degraded network using $a_{\text{IDS}}(t)$. It then evaluates candidate purification masks and selects the mask that maximizes the resource-aware score. The selected IDS-aware mask is denoted $m_{\text{IDS}}(t)$. This mask can change over time. During benign periods, it may remain close to the clean mask. During attack periods, it may become more purification-heavy in order to preserve end-to-end fidelity.

*d. Oracle-aware control*: In the oracle-aware scenario, the physical network is degraded according to $a_{\text{oracle}}(t)$, and the controller also plans using $a_{\text{oracle}}(t)$. The selected oracle-aware mask is denoted $m_{\text{oracle}}(t)$.

This is not a deployable controller because it uses ground-truth attack labels. Instead, it provides a referencfor how well the IDS-aware controller performs relative to a controller with perfect attack-rate information. For each scenario, the event-layer quantum-network simulator runs Monte Carlo trials and records the request outcome. Each trial is classified as one of the following:

- delivered above target,
- delivered below target,
- purification failed,
- swapping failed,
- generation failed, if applicable.

The main metric is above-target delivery probability,

$$U = \Pr\left(D = 1 \text{ and } F_{\text{end}} \geq F_{\star}\right),$$

where $D = 1$ means that the end-to-end entangled pair was delivered and $F_{\text{end}}$ is the final delivered fidelity. We also report raw delivery probability,

$$P_{\text{del}} = \Pr\left(D = 1\right),$$

mean delivered fidelity, $\bar{F}_{\text{del}}$, mean delivered latency, $\bar{L}_{\text{del}}$, and mean purification count, $\bar{N}_{\text{pur}}$. These metrics distinguish between delivering many low-fidelity pairs and delivering fewer but fidelity-qualified pairs.

**Attack-period-only results:** The full benign-to-SSDP trace contains both benign time bins and attack time bins. To isolate the effect of the SSDP attack, we compute a separate attack-period-only summary. A time bin is included in the attack-period-only analysis if

$$a_{\text{oracle}}(t) > 0.5.$$

This selects bins in which the majority of flows are attack traffic. For each scenario, we pool all Monte Carlo trials from the selected attack-period bins and compute: $P_{\text{del}}$, the raw delivery probability; $U$, the above-target delivery probability; $\bar{F}_{\text{del}},.$ the mean delivered fidelity; $\bar{L}_{\text{del}},.$ the mean delivered latency; and s $\bar{N}_{\text{pur}},.$ the mean purification count.

For Bernoulli metrics such as $P_{\text{del}}.$ and $U$ . , the 95% confidence half-width is computed as

$$\Delta_{95} = 1.96\sqrt{\frac{\hat{p}\left(1-\hat{p}\right)}{n}},$$

where $\hat{p}$ is the empirical probability and $n$ is the number of Monte Carlo trials. The attack-period-only analysis is important because the full-trace average includes bign bins before the SSDP transition. Tse benign bins can mask the te size of the attack effect. The attack-period-only summary therefore provides the clearest test of whether IDS-aware control improves quantum-network operation under cyber degradation.

In the attack-period-only results, the attack-unaware controller maintains high raw delivery probability but suffers a large drop in above-target delivery probability. The IDS-aware controller reduces raw delivery probability because it selects more purification-heavy masks, but it substantially increases above-target delivery probability. This demonstrates that IDS-aware control shifts the network from high-throughput low-fidelity delivery toward lower-throughput fidelity-qualified delivery. The key attack-period result is

$$U_{\text{unaware}} = 0.098 \pm 0.007,$$

for the attack-unaware controller, compared with

$$U_{\text{IDS}} = 0.344 \pm 0.011,$$

for the IDS-aware controller. The oracle-aware controller gives

$$U_{\text{oracle}} = 0.335 \pm 0.011.$$

Thus, the IDS-aware controller recovers a large fraction of the useful entanglement delivery available to the oracle-aware controller.

**Mask-adaptation mechanism:** The mask-adaptation analysis explains why IDS-aware control improves above-target delivery. Each purification policy is represented by a binary mask

$$m(t) = \left(m_1(t), m_2(t), \ldots, m_L(t)\right).$$

Here,

$$m_i(t) = 1$$

means that elementary link $i$ is purified in time bin $t$, and

$$m_i(t) = 0$$

means that it is not purified. The total number of planned purifications in time bin $t$ is

$$\left|m(t)\right| = \sum_{i=1}^{L} m_i(t).$$

In the attack-unaware scenario, the mask is fixed at the clean value:

$$m_{\text{unaware}}(t) = m_{\text{clean}}.$$

Therefore, the attack-unaware controller does not increase purification effort during the SSDP attack. It continues to deliver many pairs, but their final fidelities often fall below the target. In the IDS-aware scenario, the mask is recomputed using the IDS severity score:

$$m_{\text{IDS}}(t) = \arg\max_{m} J_t\left(m; a_{\text{IDS}}(t)\right),$$

where $J_t$ is the time-bin-specific resource-aware planning score. It depends on the estimated link degradation implied by the IDS score.

During benign time bins, $a_{\text{IDS}}(t)$ is small, so the IDS-aware mask often resembles the clean mask. During SSDP attack bins, $a_{\text{IDS}}(t)$ becomes large, and the selected mask becomes more purification-heavy. The oracle-aware scenario follows the same structure, but uses $a_{\text{oracle}}(t)$ instead of $a_{\text{IDS}}(t)$. The observed mask adaptation shows that the IDS signal changes the quantum-network control action. In the attack period, the attack-unaware controller keeps the same low-purification mask, while the IDS-aware and oracle-aware controllers switch to masks with more planned purifications.

This mechanism explains the performance trade-off:

- attack-unaware control keeps raw delivery high but loses final fidelity;
- IDS-aware control increases purification effort, lowering raw delivery but increasing above-target delivery;
- oracle-aware control provides a reference for the best performance available with perfect attack-rate information.

The key conclusion is that the IDS signal is not merely correlated with network degradation. It is used as an input to the purification planner, changes the selected purification mask, and improves the useful quantum-network outcome.

#### 1. *Experimental protocol*

The experimental protocol consisted of three linked stages. First, stationary adaptive-purification experiments were performed on three-link and eight-node linear repeater chains. CUDA-Q noisy quantum kernels were used to validate Bell-pair preparation, purification, and entanglement-swapping primitives, and the resulting success-probability and fidelity estimates were cached for use in the event-layer simulations [11, 28-31]. A SeQUeNCe-style Monte Carlo simulator then sampled elementary link generation, memory waiting-time decay, purification success or failure, entanglement swapping, final delivered fidelity, and latency for each purification policy. Second, the cyber and quantum-network components were coupled using a CSE-CIC-IDS2018 benign-to-SSDP trace [32, 33]. The oracle attack rate was used to impose time-dependent degradation on selected repeater links, while the IDS anomaly score was used by the controller to choose purification masks. Clean, attack-unaware, IDS-aware, and oracle-aware scenarios were evaluated over identical Monte Carlo trial budgets, and performance was reported using raw delivery probability, above-target delivery probability, delivered fidelity, latency, purification count, and failure-stage statistics. Uncertainties on Bernoulli metrics were reported as 95% confidence half-widths over Monte Carlo trials.

## IV. RESULTS

### A. Part 1: Adaptive purification under stationary quantum-network conditions

**CUDA-Q primitive validation:** The CUDA-Q primitive layer was validated before coupling the primitive estimates to the network-level Monte Carlo simulation. Bell-basis calibration of the entanglement-swapping estimator produced an ideal mapping accuracy of unity. For ideal inputs, the noisy CUDA-Q estimator gave an entanglement-swapping fidelity of approximately 0.991 and a purification success probability and output fidelity of approximately 0.998 and 0.997, respectively. For representative nonideal inputs, the purification estimator at elementary-pair fidelity 0.86 yielded an output fidelity near 0.889, while a swap between links of fidelity 0.91 and 0.89 yielded an end-to-end fidelity estimate near 0.810. These checks show that the CUDA-Q kernels preserve the expected ideal behavior while introducing physically meaningful degradation under noisy inputs; detailed validation results are reported in Appendix A, Table A1.

*Three-link stationary baseline:* Fig. 3(a) shows the rate-fidelity tradeoff in the three-link baseline. No purification maximizes raw delivery, with delivery probability (0.900), but none of the delivered pairs exceed the target fidelity. Threshold-adaptive purification lowers raw delivery to (0.750) and increases above-target delivery to only (0.100), indicating that a local fidelity trigger is insufficient to optimize the end-to-end chain. *Predictive adaptive purification* gives the best useful-delivery performance, increasing the above-target delivery probability to (0.684) while retaining a raw delivery probability of (0.691). *Fixed purification* produces higher conditional delivered fidelity, but its additional purification failures reduce raw delivery to (0.578) and above-target delivery to (0.578).

Thus, in the three-link setting, predictive adaptive purification identifies a favorable intermediate operating point: it purifies enough links to satisfy the fidelity target, but avoids the full failure and latency cost of purifying every elementary link.

In the 3-link stationary chain, no purification gives the highest raw delivery probability but fails the target-fidelity constraint. Fixed purification gives high fidelity but lower delivery and higher latency. The predictive adaptive policy improves the useful delivery tradeoff: it achieves above-target probability (0.684), compared with (0.578) for fixed purification, while reducing mean latency and mean purification count. This *confirms the value of optimizing end-to-end useful delivery rather than local fidelity or raw delivery alone*. The predictive adaptive policy improves above-target probability over fixed purification by $0.1059 \pm 0.0094$, a statistically meaningful Monte Carlo difference. It also improves raw delivery and reduces latency and purification overhead relative to fixed purification (Fig. 3 and Table A1 of Appendix)

*Eight-node stationary risk-aware purification:* Fig. 3(b) extends the policy comparison to the eight-node chain, where seven elementary links and six swapping operations make the end-to-end fidelity constraint more difficult to satisfy. No purification again fails the useful-delivery objective, giving essentially zero above-target probability. Mean-field predictive purification improves raw delivery but under-purifies the chain, reaching only ($0.067 \pm 0.009$) above-target delivery. Threshold-adaptive purification performs better, reaching ($0.176 \pm 0.014$), but remains limited by its local decision rule. The risk-aware predictive policy gives the strongest eight-node result, increasing above-target delivery to ($0.362 \pm 0.017$). Fixed purification reaches ($0.311 \pm 0.017$), lower than the risk-aware policy despite applying more purification. This shows that useful delivery in the longer chain is not maximized by purifying every link. Instead, the best operating point balances fidelity recovery against purification failure, latency, and resource cost.

The 8-node chain is a more demanding operating regime. No purification again gives high raw delivery but insufficient fidelity. Fixed purification improves fidelity but incurs substantial purification failure. A resource-penalized risk-aware policy chooses a partial purification mask rather than purifying every link. In the stationary 8-node experiment, the resource-aware policy achieves above-target delivery (0.362), compared with (0.311) for fixed purification, while improving raw delivery, reducing mean latency, and reducing purification count. The mechanism is that the risk-aware policy avoids unnecessary purification failures while maintaining enough fidelity margin to exceed the target.

As reflected in Table I below, Fig. 4 shows a **nonmonotonic purification tradeoff**: adding purification helps up to a point, but purifying every link is not optimal because it adds failure modes and latency. The resource-aware predictive policy achieves the largest above-target delivery probability while using fewer purification operations than fixed purification, demonstrating that *useful entanglement delivery can be improved by avoiding over-purification rather than purifying every elementary link.*

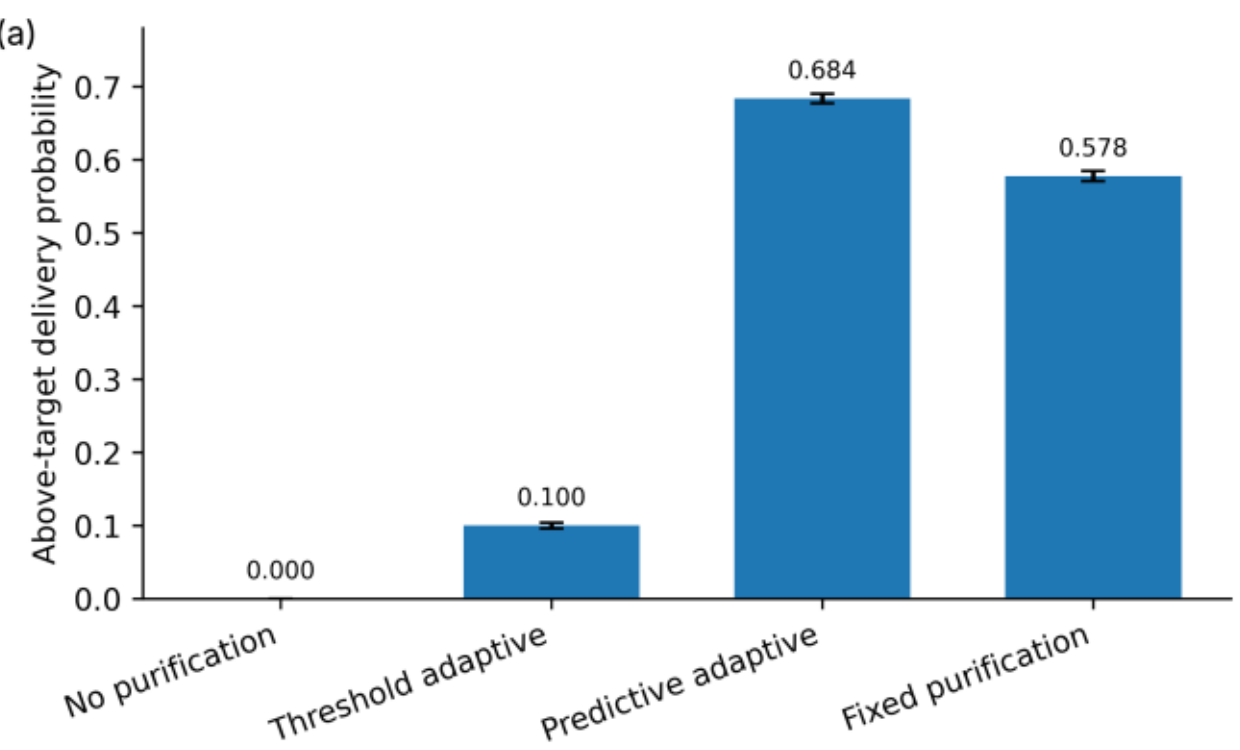

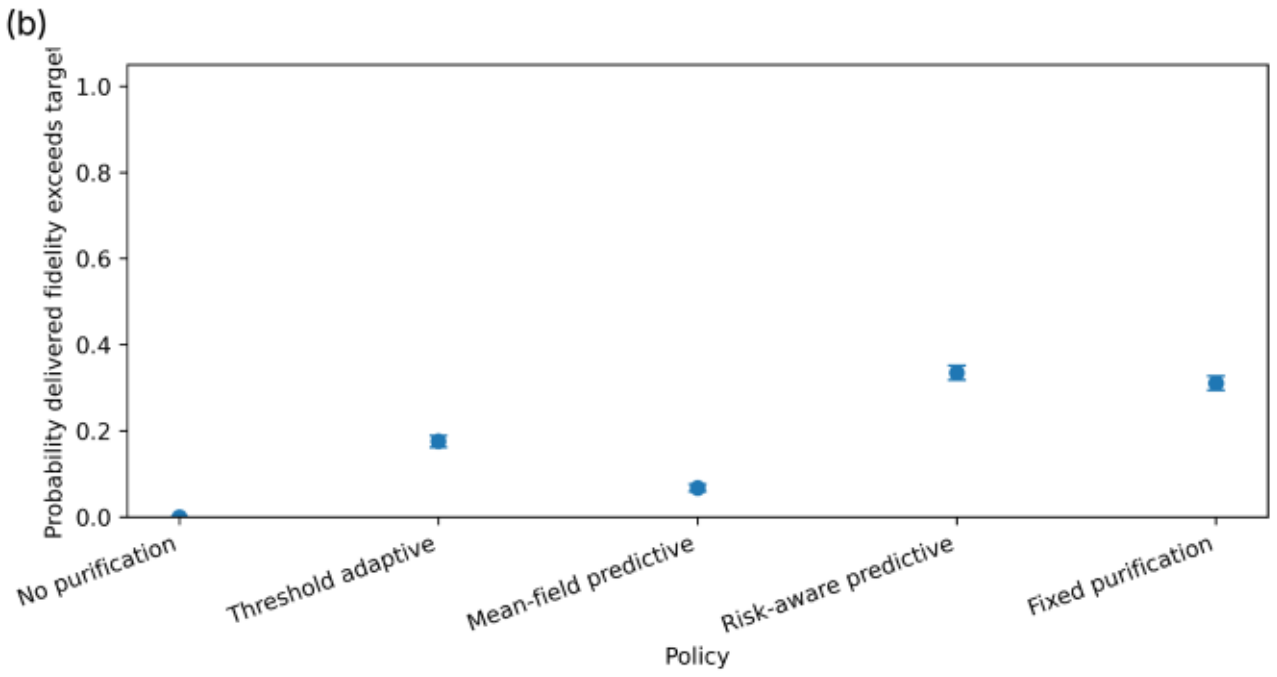

FIG. 3. (Colour online) *Stationary adaptive-purification performance*. 3(a): Three-link above-target delivery probability for four purification policies. The plot compares no purification, threshold-adaptive purification, predictive adaptive purification, and fixed purification in the three-link baseline. No purification produces no above-target deliveries at the selected target fidelity, while threshold adaptation gives only a modest useful-delivery probability. Predictive adaptive purification achieves the largest above-target delivery probability, 0.684, exceeding fixed purification, 0.578, by avoiding unnecessary purification on every elementary link. Error bars denote 95% confidence half-widths over *Monte Carlo* trials. 3(b): *Eight-node stationary policy comparison*. Eight-node above-target delivery probability for no purification, threshold-adaptive purification, mean-field predictive purification, risk-aware predictive purification, and fixed purification. The risk-aware predictive policy gives the largest above-target delivery probability, showing that useful entanglement delivery is improved by balancing fidelity gain against purification failure and resource cost. Error bars denote 95% confidence half-widths over Monte Carlo trials. Detailed results are reported in Table I.

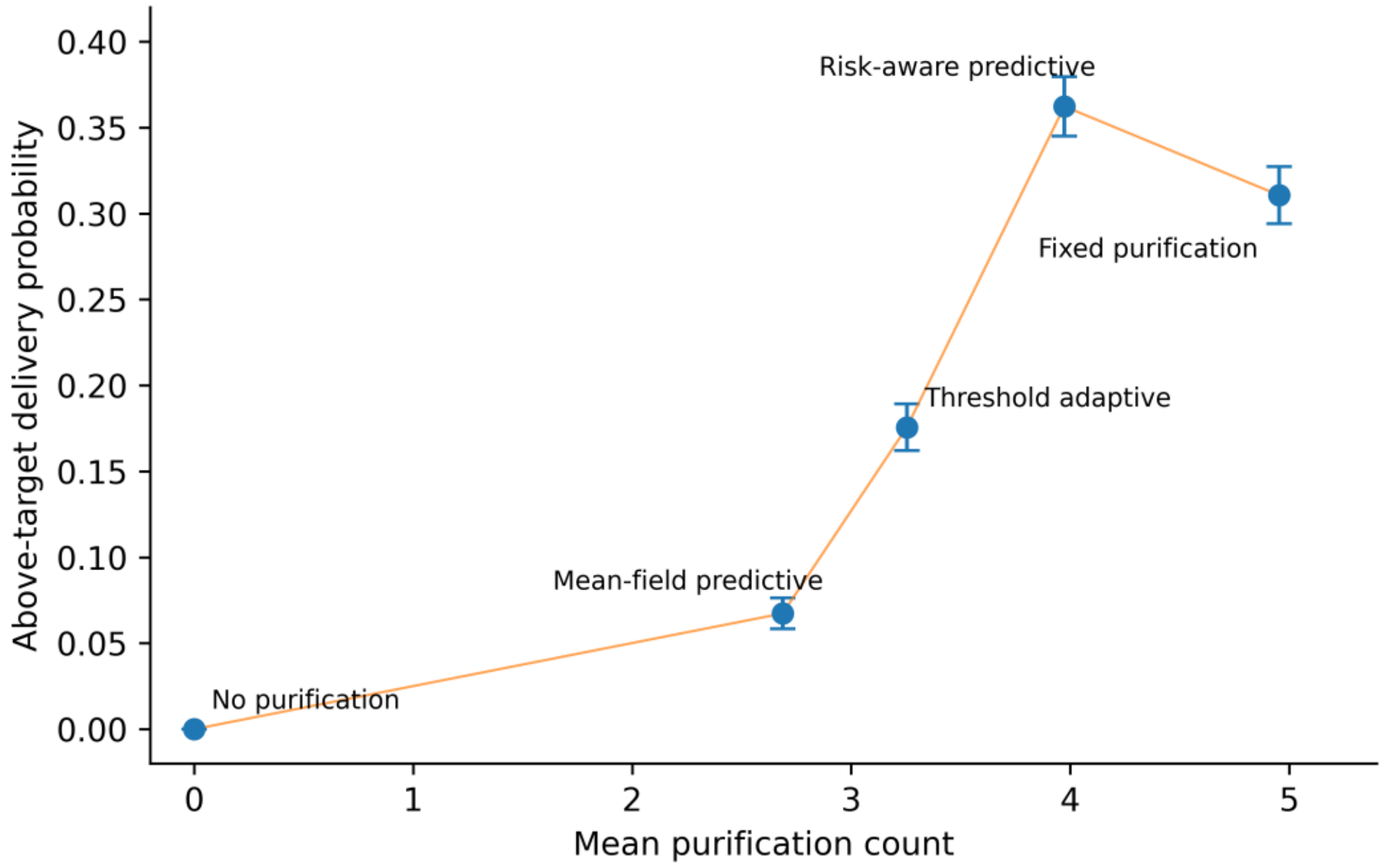

FIG. 4. (Colour online) Eight-node risk-aware purification tradeoff under stationary network conditions. The plot shows above-target delivery probability as a function of mean purification count for no purification, threshold-adaptive purification, mean-field predictive purification, resource-aware predictive purification, and fixed purification. Error bars denote 95% confidence half-widths over Monte Carlo trials.

TABLE I. Three-link *Monte Carlo* policy comparison. The table reports raw delivery probability, above-target delivery probability, mean delivered fidelity, mean delivered latency, and mean purification count for the four purification policies in the three-link baseline. Uncertainties on delivery and above-target probabilities denote 95% confidence half-widths over Monte Carlo trials. No purification gives the largest raw delivery probability but produces no above-target deliveries at the chosen fidelity threshold. Predictive adaptive purification gives the largest above-target delivery probability while requiring fewer purification operations and lower latency than fixed purification.

| Policy | Delivery | Above-target | Fidelity | Latency | Purifications |
|---|---|---|---|---|---|
| No purification | $0.8998 \pm 0.0042$ | 0 | 0.7043 | 0.997 ms | 0 |
| Threshold adaptive | $0.7496 \pm 0.0060$ | $0.1000 \pm 0.0042$ | 0.7273 | 1.391 ms | 1 |
| Predictive adaptive | $0.6914 \pm 0.0064$ | $0.6836 \pm 0.0064$ | 0.7394 | 1.494 ms | 1.886 |
| Fixed purification | $0.5777 \pm 0.0068$ | $0.5777 \pm 0.0068$ | 0.764 | 1.889 ms | 2.628 |

The eight-node sweep results in Appendix A, Fig10 further show that the policy comparison is operating-regime dependent. Local threshold adaptation is sensitive to the chosen purification trigger, while target-fidelity sweeps reveal a narrow intermediate regime in which adaptive purification can improve useful delivery before the end-to-end fidelity requirement becomes too stringent for all policies. The supporting diagnostics show that the risk-aware policy improves useful delivery by remaining above the fidelity threshold while avoiding the raw-delivery and latency penalties of fixed purification.

Table I shows the basic rate-fidelity-resource tradeoff in the three-link chain. The no-purification policy maximizes raw delivery probability, $0.8998 \pm 0.0042$, because it avoids purification failures and extra latency. However, its mean delivered fidelity is only $0.7043$, so none of the delivered pairs satisfy the target-fidelity condition. Threshold-adaptive purification improves the mean delivered fidelity to $0.7273$ but its above-target delivery probability remains low, $0.1000 \pm 0.0042$, indicating that a local threshold rule is not sufficient to optimize useful end-to-end delivery.

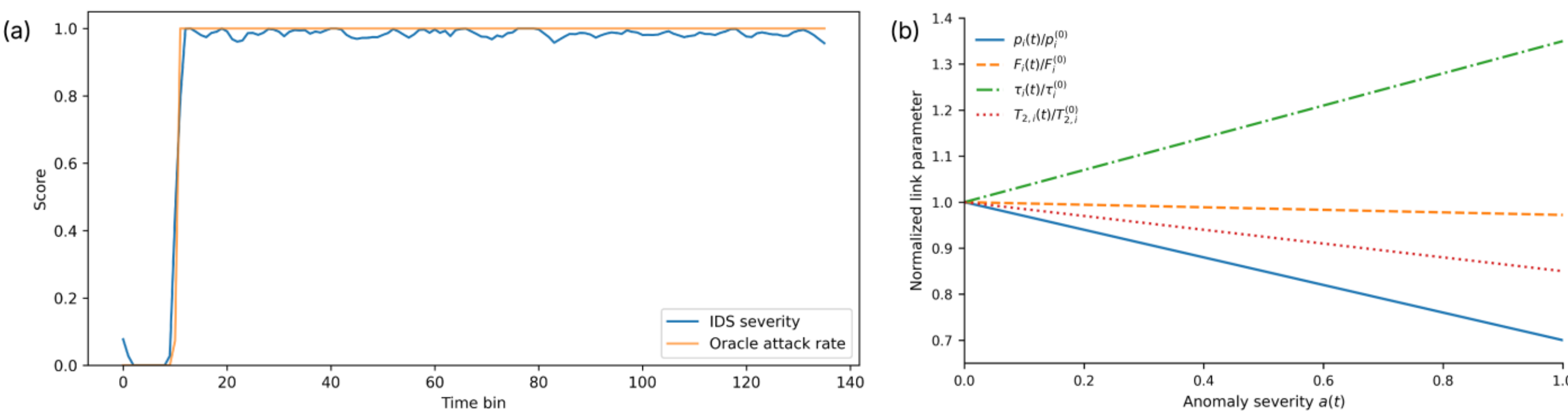

(c)

| Scenario | Physical links | Controller input | Mask used | Purpose |
|---|---|---|---|---|
| Clean counterfactual | Baseline links (no degradation) | Clean network parameters | $m_{\mathrm{clean}}$ | No-attack reference |
| Attack unaware | Degraded by oracle attack rate | Assumes clean conditions | $m_{\mathrm{clean}}$ | Stationary control under attack |
| IDS aware | Degraded by oracle attack rate | IDS severity\na$_{\mathrm{IDS}}(t)$ | $m_{\mathrm{IDS}}(t)$ | Deployable cyber-aware control |
| Oracle aware | Degraded by oracle attack rate | Oracle rate\na$_{\mathrm{oracle}}(t)$ | $m_{\mathrm{oracle}}(t)$ | Upper-reference controller |

FIG 5. (Colour online). IDS severity and oracle attack rate across time bins constructed from CSE-CIC-IDS2018 benign and SSDP traffic. 5(a) The IDS score rises near the attack transition and remains high during the SSDP period. 5(b). Cyber-to-quantum degradation model used in the IDS-aware repeater simulations. The anomaly severity ( $a(t) \in [0,1]$ ) modifies selected elementary-link parameters by reducing the generation probability ( $p_i(t)$ ), reducing the raw fidelity ( $F_i(t)$ ), increasing the attempt time ($\tau_i(t)$), and reducing the memory coherence time ( $T_{2,i}(t)$ ). The plotted curves show the normalized parameter changes used to map classical attack severity into time-dependent quantum-network degradation. 5(c). Controller scenarios used in the coupled cyber–quantum experiment. The clean counterfactual uses baseline link parameters without degradation. The attack-unaware controller experiences oracle-defined physical degradation but continues to use the clean purification mask. The IDS-aware controller experiences the same physical degradation but selects time-dependent masks from the learned IDS severity signal. The oracle-aware controller uses the ground-truth attack rate and serves as an upper-reference controller.

In Table I, Predictive adaptive purification gives the strongest useful-delivery result. It achieves an above-target delivery probability of $0.6836 \pm 0.0064$, substantially higher than fixed purification, $0.5777 \pm 0.0068$, while also improving raw delivery probability and reducing latency. Compared with fixed purification, the predictive policy increases above-target delivery by $0.1059$, reduces mean latency from $1.889$ ms to $1.494$ ms, and lowers the mean purification count from $2.628$ to $1.886$. Fixed purification gives the highest mean delivered fidelity, $0.7640$, but its extra purification attempts introduce additional failure modes and latency. Thus, the three-link baseline demonstrates that maximizing delivered fidelity alone is not the same as maximizing useful entanglement delivery; the best operating point is obtained by selectively purifying the links needed to exceed the target fidelity.

TABLE II. 8-node stationary tradeoff between purification cost and above-target delivery probability

| Policy | Mean purifications | Above-target delivery probability |
|---|---|---|
| No purification | 0 | 0 |
| Mean-field adaptive | 2.687 | $0.067 \pm 0.009$ |
| Threshold adaptive | 3.254 | $0.176 \pm 0.014$ |
| Risk-aware predictive | 3.973 | $0.362 \pm 0.017$ |
| Fixed purification | 4.954 | $0.311 \pm 0.017$ |

Table II reports the mean number of purification operations per trial and the probability of delivering an end-to-end entangled pair above the target fidelity for each policy. No purification produces no above-target deliveries, while mean-field predictive and threshold-adaptive purification improve the useful-delivery probability only modestly. The risk-aware predictive policy achieves the largest above-target delivery probability, ($0.362 \pm 0.017$), while using fewer purifications than fixed purification. Uncertainties denote 95% confidence half-widths over Monte Carlo trials.

### B. Part2: IDS-aware SSDP experiment

**IDS-aware experiment:** This section links the CSE-CIC-IDS2018 benign-to-SSDP anomaly trace to quantum-network degradation. During benign bins, the clean, attack-unaware, IDS-aware, and oracle-aware scenarios remain similar. Once the SSDP attack begins, the attack-unaware controller keeps its clean purification mask and still delivers many pairs, but most fall below the target fidelity. In contrast, the IDS-aware controller detects high anomaly severity and switches to more purification-heavy masks, reducing raw delivery but increasing fidelity-qualified delivery.

Across the full trace, above-target delivery rises from $0.291$ for the attack-unaware controller to $0.447$ for the IDS-aware controller. During the attack-only period, the gain is larger: attack-unaware delivery is $0.098$, while IDS-aware delivery is $0.344$, close to the oracle-aware value of $0.335$. Thus, the IDS-derived anomaly score recovers much of the useful entanglement delivery lost under attack.

Fig. 5 presents the coupled cyber–quantum control experiment. Fig. 5(a) shows the benign-to-SSDP trace built from CSE-CIC-IDS2018 data: the oracle attack-rate signal stays low during benign traffic and rises sharply when SSDP dominates the flow bins. The IDS-derived severity follows this transition without using ground-truth labels during control, providing a time-dependent cyber-state signal for the purification controller while retaining the oracle trace as a reference for physical degradation and upper-bound control.

Fig. 5(b) shows the phenomenological cyber-to-quantum degradation model that maps attack severity to repeater-link degradation. As anomaly severity $a(t)$ increases, elementary link-generation probability $p_i(t)$, raw Bell-pair fidelity $F_i(t)$, and memory coherence time $T_{2,i}(t)$ decrease, while effective attempt time $\tau_i(t)$ increases. The attack therefore does not directly alter the quantum state; instead, it degrades the operating conditions for generating, storing, and connecting elementary entanglement. This mapping lets the classical cyber trace enter the quantum-network simulation through physically interpretable link parameters.

Fig. 5(c) summarizes the four controller scenarios used in the coupled experiment. The clean counterfactual evaluates the repeater chain without cyber degradation. The attack-unaware controller experiences attack-induced degradation but continues to use the clean purification mask. The IDS-aware controller experiences the same physical degradation but replans the purification mask using the IDS severity signal. Finally, the oracle-aware controller replans using the ground-truth attack rate and therefore provides a reference for the best performance obtainable with perfect cyber-state information. The comparison in Fig 5(c) separates three effects: degradation of the physical network, availability of cyber-state information, and the quality of the cyber-state estimate used for control.

Fig. 6 reports the main IDS-aware control result during the SSDP attack period. Fig. 6(a) shows that the attack-unaware controller suffers a strong reduction in above-target entanglement delivery, reaching only ($0.098 \pm 0.007$). In contrast, IDS-aware control increases the above-target delivery probability to ($0.344 \pm 0.011$), close to the oracle-aware value of ($0.335 \pm 0.011$). The comparison shows that the IDS severity signal is sufficiently informative to recover much of the useful entanglement delivery that would otherwise be lost under attack-induced degradation.

Fig. 6(b) separates raw entanglement delivery from fidelity-qualified delivery during the SSDP attack period. The attack-unaware controller continues to use the clean purification mask, so it preserves a high raw delivery probability. However, most of those delivered pairs fall below the target fidelity, producing a low above-target delivery probability of ($0.098 \pm 0.007$). IDS-aware control changes this operating point. By using the IDS severity signal to select more purification-heavy masks, it sacrifices raw delivery probability but recovers useful entanglement delivery, increasing the above-target probability to ($0.344 \pm 0.011$). The oracle-aware controller follows the qualitative behavior and remains close to the IDS-aware result, with raw delivery probability approximately (0.346) and above-target delivery probability approximately (0.335), indicating that the learned IDS severity provides nearly the same control-relevant information as the ground-truth attack rate. This figure shows that the advantage of IDS-aware control is not higher throughput in the raw sense, but a shift from high-volume low-fidelity delivery to lower-volume fidelity-qualified delivery. The IDS-aware control trades raw throughput for fidelity-qualified throughput, which is the relevant objective for useful entanglement distribution.

Fig. 6(c) shows that the performance gain is accompanied by a change in the selected purification masks. The attack-unaware controller keeps the clean mask fixed throughout the trace and therefore does not respond to the SSDP transition. By contrast, the IDS-aware controller switches during the attack period to masks with more planned purifications, closely matching the behavior of the oracle-aware controller.

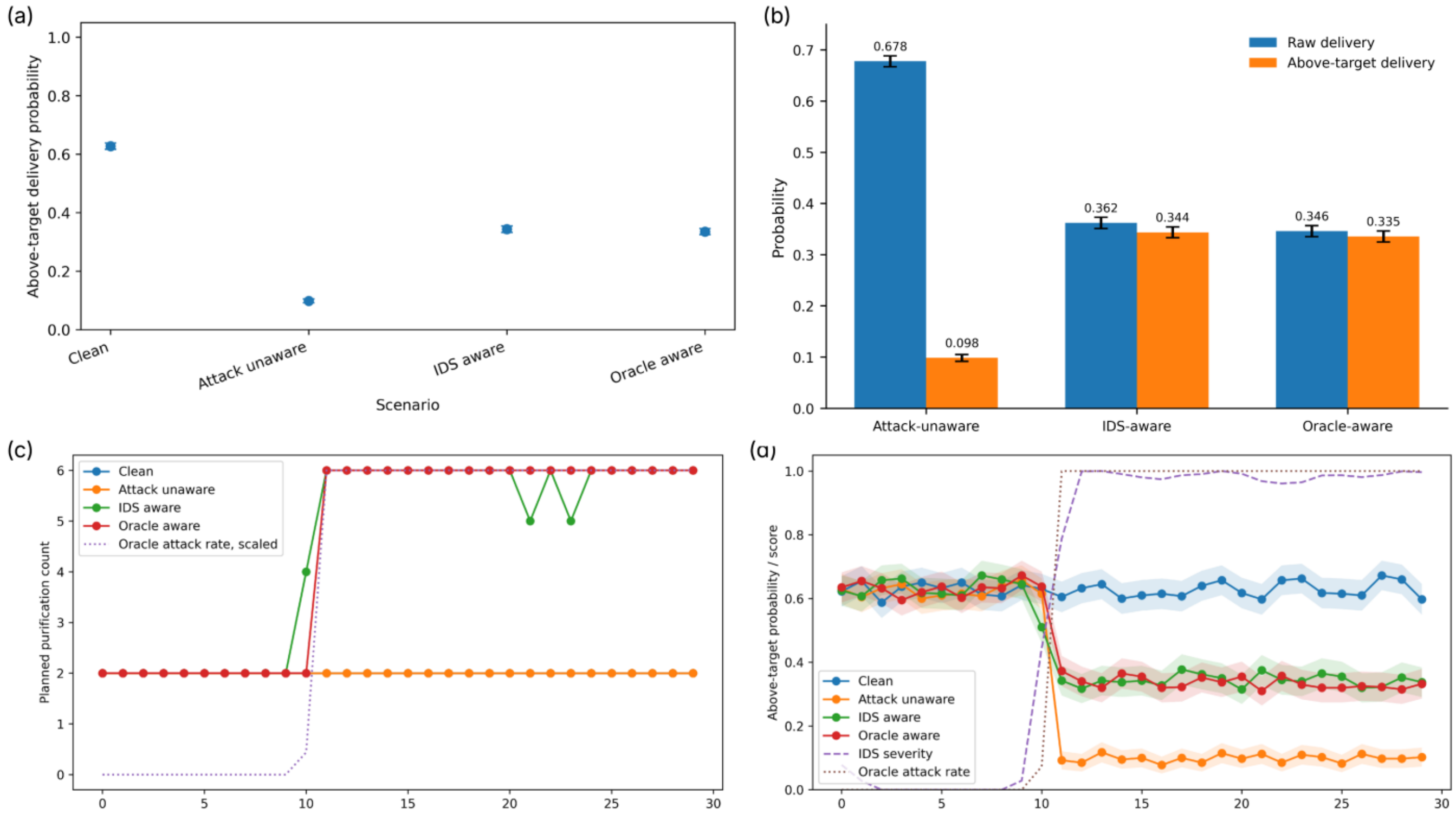

FIG. 6. (Colour online). Attack-period above-target entanglement-delivery probability for attack-unaware, IDS-aware, and oracle-aware controllers. 6(a) The attack-unaware controller delivers many below-target pairs under cyber-induced degradation, giving an above-target probability of ( $0.098 \pm 0.007$ ). IDS-aware control increases above-target delivery to ( $0.344 \pm 0.011$ ), close to the oracle-aware value of ( $0.335 \pm 0.011$ ). Error bars denote 95% confidence half-widths over Monte Carlo trials. 6(b). Attack-period tradeoff between raw delivery and above-target delivery under SSDP degradation. The attack-unaware controller maintains a high raw delivery probability, ( $0.678 \pm 0.011$ ), but only ( $0.098 \pm 0.007$ ) of trials deliver entanglement above the target fidelity. IDS-aware control lowers raw delivery to ( $0.362 \pm 0.011$ ) because it selects more purification-heavy masks, but increases above-target delivery to ( $0.344 \pm 0.011$ ). The oracle-aware controller shows a similar tradeoff, with raw delivery ( $0.346 \pm 0.011$ ) and above-target delivery ( $0.335 \pm 0.011$ ). Error bars denote 95% confidence half-widths over *Monte Carlo trials*. 6(c). Purification-mask adaptation during the benign-to-SSDP trace. The attack-unaware controller keeps the clean purification mask fixed under attack, while the IDS-aware controller changes to more purification-heavy masks as the IDS severity rises. The oracle-aware controller provides a reference based on ground-truth attack-rate information. The mask transition shows that cyber-state information changes the selected purification action, leading to improved fidelity-qualified entanglement delivery. 6(d). *Time-resolved* above-target delivery probability during the benign-to-SSDP transition. The above-target delivery probability versus time bin plot shows, the clean counterfactual remains near the stationary no-attack performance throughout the trace. After the oracle attack rate and IDS severity rise, the attack-unaware controller collapses to a low above-target delivery probability, while the IDS-aware and oracle-aware controllers maintain substantially higher fidelity-qualified delivery by adapting the purification policy. Shaded bands denote 95% confidence intervals over Monte Carlo trials in each time bin.

In the final attack-period regime in Fig. 6(d), the IDS-aware and oracle-aware policies most frequently select higher-purification masks such as 1111011, whereas the attack-unaware controller remains at the lower-purification clean mask. This confirms that the IDS signal is not only correlated with the degradation; it actively changes the quantum-network control action and thereby improves fidelity-qualified delivery. The time-resolved control traces support the attack-period summary in Fig. 6(c). Before the SSDP transition, all degraded-network scenarios are close to the clean counterfactual, with above-target delivery probabilities near $0.6$. When the oracle attack rate rises sharply and the IDS severity approaches unity, the attack-unaware controller experiences a rapid collapse in above-target delivery, falling to approximately $0.08-0.12$. By contrast, the IDS-aware and oracle-aware controllers decrease less severely and stabilize near $0.3-0.4$. This behavior shows that the IDS-aware controller does not merely improve the attack-period average; it dynamically responds to the cyber-state transition and preserves fidelity-qualified delivery throughout the degraded regime.

### *1. Discussions*

In the 8-node repeater chain, SSDP-driven degradation reduced attack-unaware above-target entanglement delivery from the clean counterfactual value of 0.627±0.011to 0.098±0.007. Incorporating IDS-derived anomaly severity into the purification planner increased above-target delivery to 0.344±0.011, close to the oracle-aware value of 0.335±0.011. This recovery came at the expected cost of lower raw delivery probability, increased purification use, and higher latency, indicating a deliberate shift from high-throughput low-fidelity delivery toward lower-throughput useful entanglement delivery.

IDS-aware resource-adaptive purification improves useful above-target entanglement delivery under SSDP-driven cyber degradation. During the attack period, the IDS-aware controller increases above-target delivery from 0.098 ± 0.007 to 0.344 ± 0.011, closely matching the oracle-aware controller, by switching to more purification-heavy masks. This gain is achieved at the expected cost of lower raw delivery and increased latency, consistent with a fidelity-constrained quantum-network service objective.

## V. CONCLUSION

We have presented a cyber-aware quantum-network simulation workflow that links adaptive purification in quantum-repeater chains with anomaly-informed control. The study was organized around two main connected questions: how purification should be allocated under stationary quantum-network conditions, and whether an IDS-derived cyber-state signal can be used to improve quantum-network control under attack-induced degradation. The resulting framework combines CUDA-Q primitive validation, SeQUeNCe-style event-layer Monte Carlo simulation, risk-aware purification planning, and IDS-aware repeater control.

In the stationary quantum-network experiments, CUDA-Q noisy kernels were used to validate Bell-pair preparation, purification, and entanglement-swapping primitives before embedding their estimated success probabilities and output fidelities into the event-layer repeater simulation. The three-link baseline showed the basic rate-fidelity-resource tradeoff. No purification preserved the highest raw delivery probability but failed the above-target fidelity requirement. Fixed purification increased delivered fidelity but introduced additional failure and latency costs. Predictive adaptive purification achieved the strongest useful-delivery performance in the three-link setting, increasing above-target delivery relative to fixed purification while using fewer purification operations. The eight-node experiments extended this result to a longer heterogeneous chain. There, the resource-aware predictive policy achieved the largest above-target delivery probability, $0.362 \pm 0.017$, while using fewer purifications than fixed purification, which achieved $0.311 \pm 0.017$. *These results show that useful entanglement delivery is not maximized by purifying every elementary link; rather, it requires balancing fidelity gain against purification failure, latency, and resource cost*.

The final coupled experiment showed that *cyber-state information can become actionable for quantum-network control*. A CSE-CIC-IDS2018 benign-to-SSDP trace was converted into an IDS severity signal and an oracle attack-rate reference. The oracle signal defined the physical degradation applied to selected repeater links, while the IDS signal was used by the deployable controller to select purification masks. During the SSDP attack period, the attack-unaware controller maintained high raw delivery but delivered many pairs below the target fidelity, reducing above-target delivery to $0.098 \pm 0.007$. IDS-aware control selected more purification-heavy masks and increased above-target delivery to $0.344 \pm 0.011$, close to the oracle-aware value of $0.335 \pm 0.011$. This *result demonstrates* that *the IDS signal is not merely correlated with degradation; it changes the purification action and improves the fidelity-qualified quantum-network outcome*.

Overall, the *main conclusion* is that adaptive purification should be treated as a *cyber-aware control* problem rather than a static link-level operation. In stationary networks, risk-aware purification improves the rate-fidelity-resource tradeoff by *avoiding unnecessary purification*. Under SSDP-induced degradation, IDS-aware control recovers useful above-target entanglement delivery by increasing purification effort only when the cyber state indicates that the network has degraded. The results therefore support a restrained but positive claim: *hybrid quantum models and CUDA-Q-based primitive estimators are most useful here not as standalone replacements for classical methods, but as components of a larger cyber-aware quantum-network control workflow*.

*Limitations*: The cyber-to-quantum degradation model is phenomenological: it represents the operational impact of a degraded classical control environment on link generation probability, raw fidelity, attempt time, and memory coherence, but it is not a hardware-calibrated model of a specific deployed quantum network. The repeater simulations use linear chains and simplified stochastic models for generation, purification, swapping, and memory decay. The CUDA-Q kernels validate primitive operations, while the full network dynamics are evaluated at the event layer rather than by simulating the entire repeater chain as one global quantum circuit. Finally, the IDS-aware experiment uses a controlled benign-to-SSDP trace and should be interpreted as a proof-of-principle demonstration of cyber-aware control rather than a complete security model for quantum-network infrastructure.

*Outlook*: Future work should extend this framework in several directions. The degradation model can be calibrated against hardware or testbed measurements of timing, synchronization, classical-control congestion, and memory performance. The network model can be expanded beyond linear chains to heterogeneous repeater graphs with routing, multiplexing, and competing requests. The adaptive controller can be improved using online learning, uncertainty-aware planning, and multi-objective optimization over throughput, fidelity, latency, memory occupation, and security state. On the cyber side, a hybrid quantum anomaly-analysis layer can be tested on broader attack mixtures, streaming traces, and hardware-executed variational circuits. These extensions would help determine when quantum latent diagnostics provide operational value beyond classical IDS outputs.

## APPENDIX A: SUPPORTING DATA

*CUDA-Q primitive-validation*: The CUDA-Q primitive layer was validated before coupling the primitive estimates to the network-level Monte Carlo simulation. Bell-basis calibration of the entanglement-swapping estimator produced an ideal mapping accuracy of unity. For ideal inputs, the noisy CUDA-Q estimator gave an entanglement-swapping fidelity of approximately $0.991$ and a purification success probability and output fidelity of approximately $0.998$ and $0.997$, respectively.

Table A1 reports the primitive-level checks used to validate the CUDA-Q component of the simulation workflow. The Bell-pair preparation test verifies the measurement convention by producing the expected even-parity Bell-state outcomes. The ideal swapping and purification tests confirm that the primitive circuits recover the correct high-fidelity limits before noise is introduced. The noisy swapping example then verifies that the estimator responds correctly to imperfect input links: when the input fidelities are $F_1 = 0.91$ and $F_2 = 0.89$, the output fidelity decreases to approximately $0.81$, consistent with the expected degradation from composing noisy elementary links. These results validate the primitive fidelity and success-probability estimates that are cached and supplied to the SeQUeNCe-style event-layer Monte Carlo model. For representative nonideal inputs, the purification estimator at elementary-pair fidelity $0.86$ yielded an output fidelity near $0.889$, while a swap between links of fidelity $0.91$ and $0.89$ yielded an end-to-end fidelity estimate near $0.810$. These checks show that the CUDA-Q kernels preserve the expected ideal behavior while introducing physically meaningful degradation under noisy inputs.

TABLE A1. Primitive-level checks for CUDAQ validation. Bell-pair preparation, entanglement swapping, and entanglement purification were validated before embedding the primitive estimates into the event-layer repeater simulation. Ideal circuits produced the expected limiting behavior: Bell-pair preparation gave approximately balanced $00$ and $11$ outcomes, while ideal swapping and ideal purification remained close to unit fidelity or success probability. Noisy inputs reduced the estimated output fidelity in the expected direction; for example, noisy swapping with $F_1 = 0.91$ and $F_2 = 0.89$ produced an output fidelity near $0.81$. These checks confirm that the CUDA-Q noisy kernels provide calibrated primitive estimators for the network-level Monte Carlo planner.

| Primitive | Input fidelities | Quantity checked | Expected behaviour | Observed result |
|---|---|---|---|---|
| Bell-pair preparation | ideal | counts | only counts $00$ and $11$ | approximately balanced $00/11$ |
| Entanglement swapping | ideal | endpoint fidelity | near 1 | near 1 |
| Purification | ideal | success probability and output fidelity | near 1 | near 1 |
| Noisy swapping | $F_1 = .91$, $F_2 = 0.89$ | output fidelity | below inputs / fidelity loss | $\approx 0.81$ |
| Noisy purification | noisy inputs | Success probability & conditional fidelity | finite success, improved conditional fidelity | calibrated estimate |

TABLE A2. Eight-node stationary policy comparison. Eight-node stationary tradeoff between purification cost and above-target delivery probability. The table reports the mean number of purification operations per trial and the probability of delivering an end-to-end entangled pair above the target fidelity for each policy. No purification produces no above-target deliveries, while mean-field predictive and threshold-adaptive purification improve the useful-delivery probability only modestly. The risk-aware predictive policy achieves the largest above-target delivery probability, $0.362 \pm 0.017$, while using fewer purifications than fixed purification. Uncertainties denote 95% confidence half-widths over Monte Carlo trials.

| Policy | Raw delivery | Above-target delivery | Mean fidelity | Mean latency(ms) | Mean attempts | Mean purifications |
|---|---|---|---|---|---|---|
| No purification | 0.829 ± 0.0042 | 0 | 0.518 | 1.788 | 15.742 | 0 |
| Threshold adaptive | 0.464 ± 0.0060 | 0.176 ± 0.0042 | 0.552 | 2.641 | 20.184 | 3.254 |
| Mean-field predictive | 0.578 ± 0.0064 | 0.067 ± 0.0064 | 0.545 | 2.487 | 19.701 | 2.687 |
| Risk-aware predictive | 0.423 ± 0.0068 | 0.362 ± 0.0068 | 0.573 | 2.925 | 21.412 | 3.973 |
| Fixed purification | 0.326 ± 0.0061 | 0.311 ± 0.0061 | 0.584 | 3.266 | 22.953 | 4.954 |

Table A2 reports raw delivery probability, above-target delivery probability, delivered fidelity, delivered latency, generation-attempt count, and purification count for each purification policy. The risk-aware predictive policy gives the highest above-target delivery probability while using fewer purifications than fixed purification, showing that useful delivery is improved by balancing fidelity recovery against purification failure and resource cost. The error bars support the above-target improvement at the Monte Carlo level. The status table confirms the mechanism: fixed purification experiences more purification failures, while the resource-aware policy avoids some of those failures by using fewer purification operations.

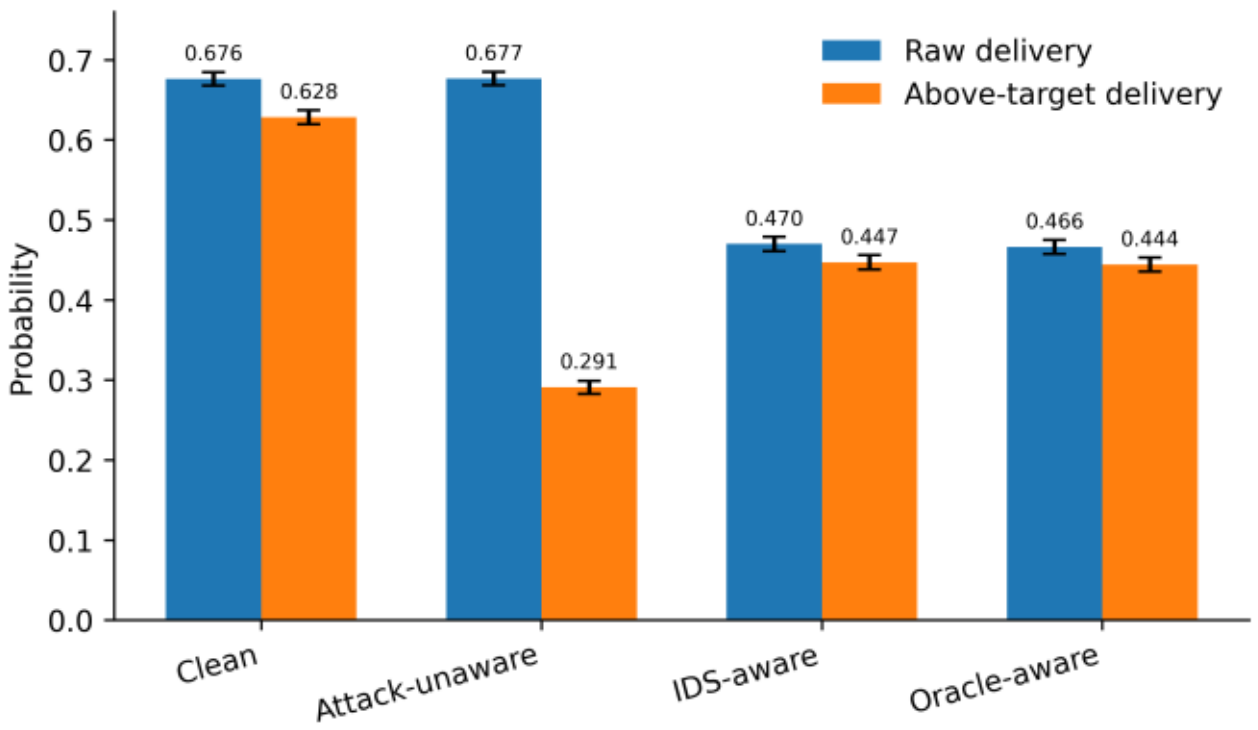

FIG. 7. (Colour online) Attack-period tradeoff between raw delivery and above-target delivery under SSDP degradation. The attack-unaware controller maintains a high raw delivery probability, $0.678 \pm 0.011$, but only $0.098 \pm 0.007$ of trials deliver entanglement above the target fidelity. IDS-aware control lowers raw delivery to $0.362 \pm 0.011$ because it selects more purification-heavy masks, but increases above-target delivery to $0.344 \pm 0.011$. The oracle-aware controller shows a similar tradeoff, with raw delivery $0.346 \pm 0.011$ and above-target delivery $0.335 \pm 0.011$. Error bars denote 95% confidence half-widths over Monte Carlo trials.

Fig. 7 separates raw entanglement delivery from fidelity-qualified delivery during the SSDP attack period. The attack-unaware controller continues to use the clean purification mask, so it preserves a high raw delivery probability. However, most of those delivered pairs fall below the target fidelities, producing a low above-target delivery probability of $0.098 \pm 0.007$. IDS-aware control changes this operating point. By using the IDS severity signal to select more purification-heavy masks, it sacrifices raw delivery probability but recovers useful entanglement delivery, increasing the above-target probability to $0.344 \pm 0.011$. The oracle-aware controller follows the same pattern and remains close to the IDS-aware result, indicating that the learned IDS severity provides nearly the same control-relevant information as the ground-truth attack rate. This figure shows that the advantage of IDS-aware control is not higher throughput in the raw sense, but a shift from high-volume low-fidelity delivery to lower-volume fidelity-qualified delivery.

Fig. 8 explains the mechanism behind the improvement in above-target delivery. The attack-unaware controller uses the same purification decision before and after the SSDP transition, so it does not compensate for the attack-induced reduction in link quality. As a result, it can maintain raw delivery but many delivered pairs fall below the target fidelity. The IDS-aware controller responds differently: once the IDS severity rises, it selects masks with a larger planned purification count. The oracle-aware controller shows similar behavior when given the ground-truth attack-rate signal. This agreement indicates that the IDS severity trace provides actionable information for the purification planner. The figure therefore supports the central control claim: IDS-aware adaptation improves useful entanglement delivery because it changes the selected purification mask, not merely because it observes the attack.

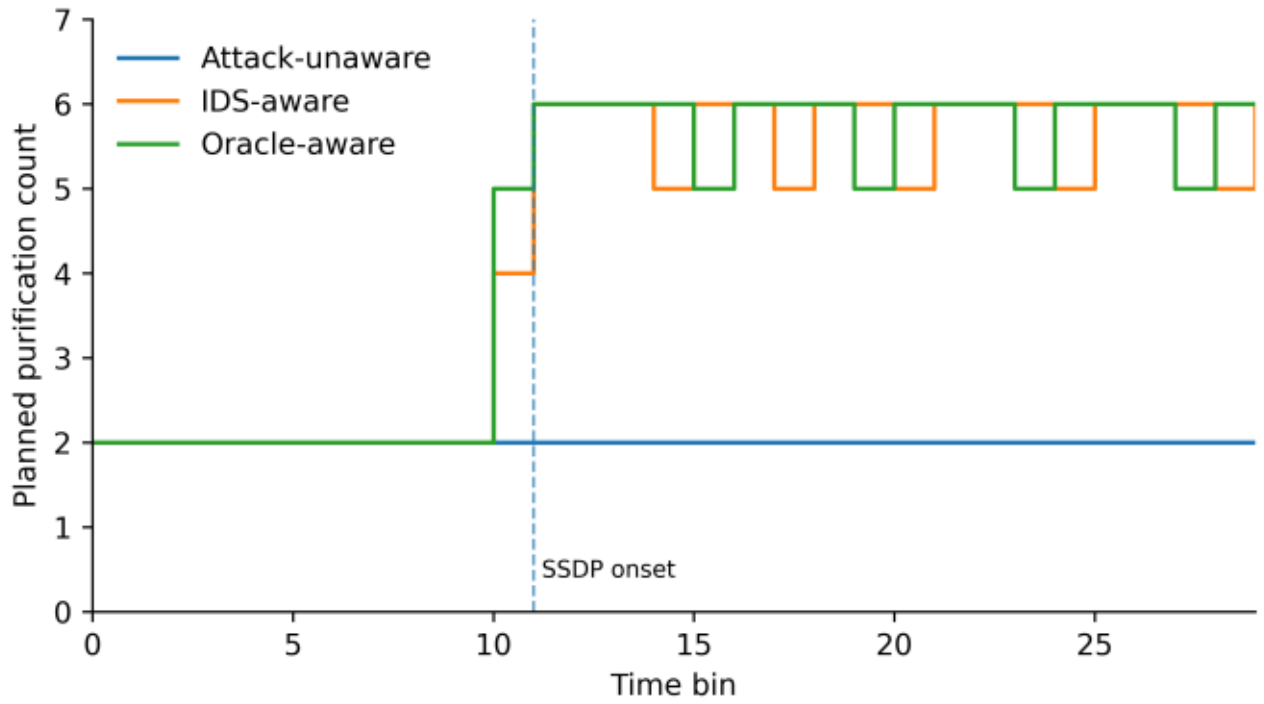

FIG. 8. (Colour online) Schematic mask adaptation under SSDP degradation. The plot shows the planned purification count as a function of time bin for the attack-unaware, IDS-aware, and oracle-aware controllers. The attack-unaware controller keeps the clean purification mask fixed during the attack period, corresponding to a constant low purification count. In contrast, the IDS-aware and oracle-aware controllers switch to more purification-heavy masks after the SSDP transition, reflecting the increased purification effort needed to preserve end-to-end fidelity under degraded link conditions. The vertical dashed line marks the attack onset. This schematic illustrates the control mechanism by which cyber-state information changes the quantum-network purification action.

Fig. 9(a) shows that the local threshold-adaptive policy is highly sensitive to the purification-trigger value. For low trigger values, the threshold rule selects too little purification and remains close to the no-purification operating point. When the trigger reaches the scale of the weaker elementary-link fidelities, the above-target delivery probability increases sharply. At still higher trigger values, the policy begins to purify too aggressively, and its performance moves back toward the fixed-purification limit. In contrast, the predictive-adaptive and fixed-purification policies are nearly independent of the trigger parameter. This sweep shows why local threshold tuning is brittle in a heterogeneous eight-node chain, whereas predictive policies are more stable because they evaluate end-to-end performance rather than only local link quality.

Fig. 9(b) shows the dependence of above-target delivery on the required target fidelity. At low target thresholds, several policies can deliver above-target entanglement with nonzero probability. As the target threshold increases, the useful-delivery probability decreases for all policies because the end-to-end fidelity must survive elementary-link noise, memory waiting-time decay, purification failure, and multiple swapping operations.

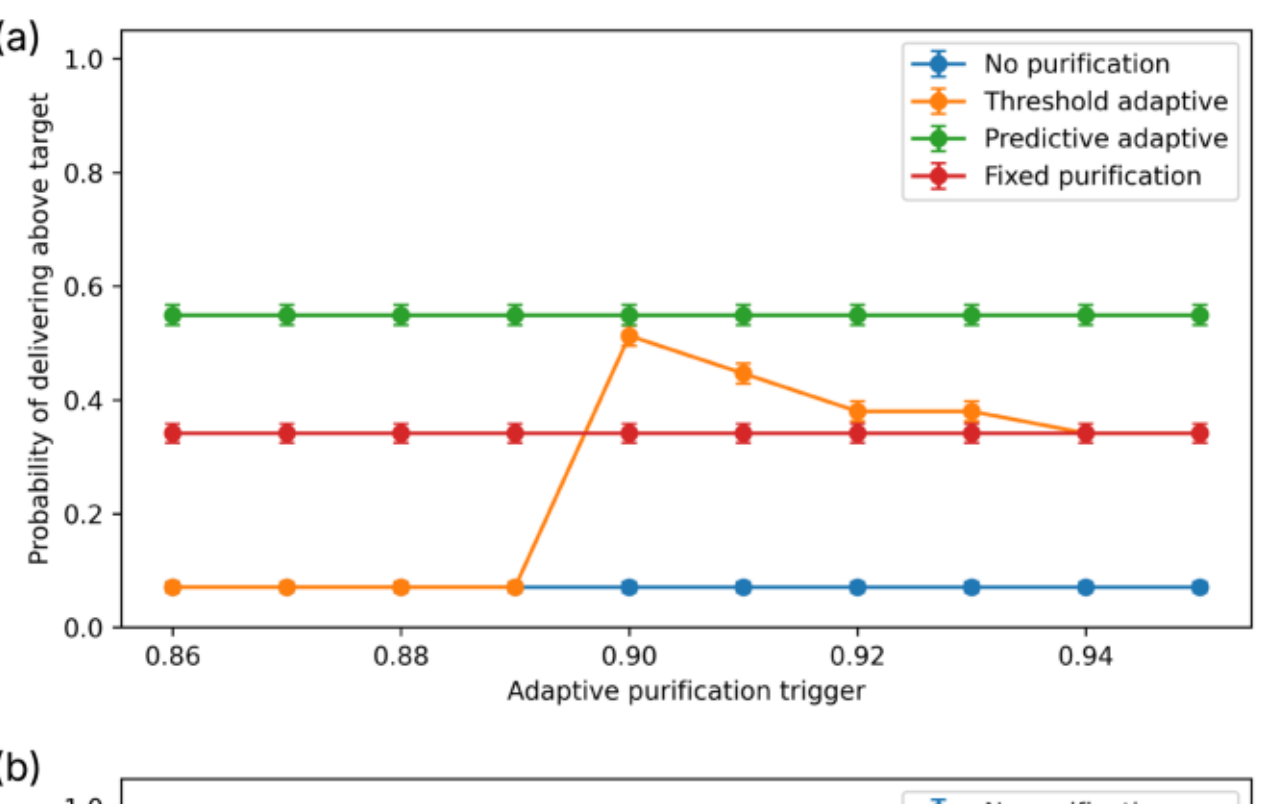

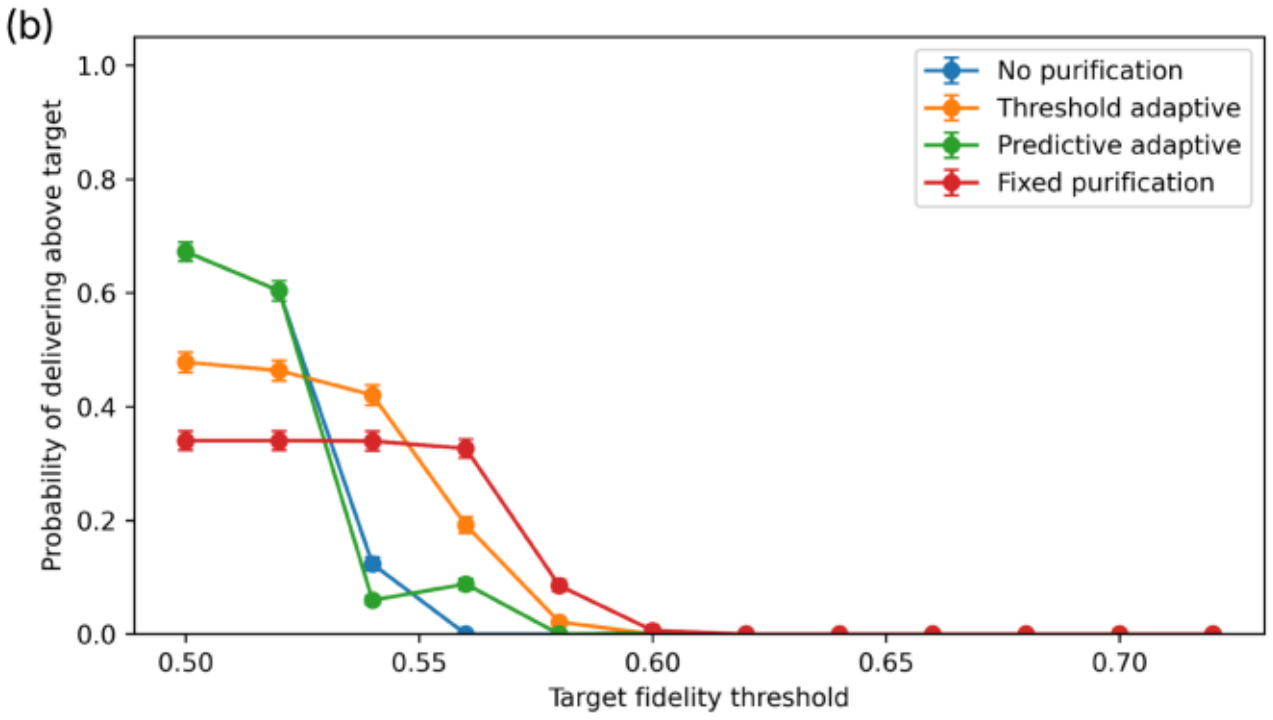

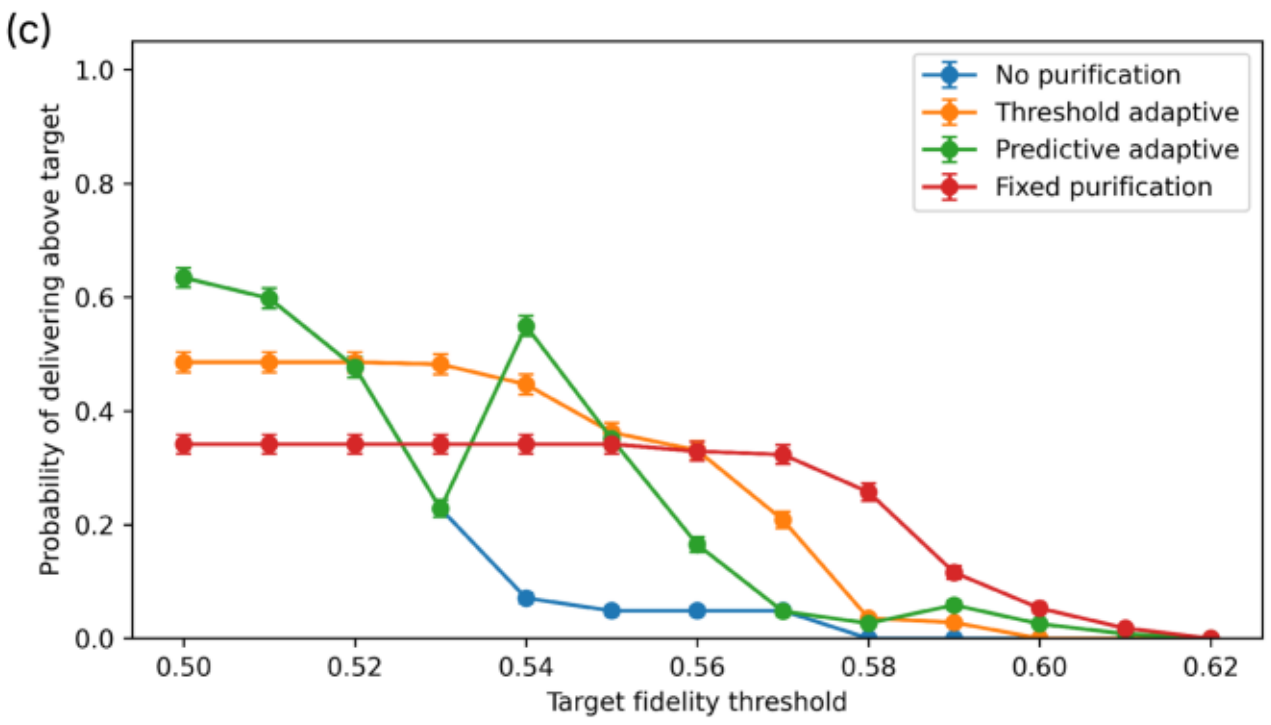

FIG. 9. (Colour online) Eight-node stationary robustness sweeps. (a) Above-target delivery probability as a function of the adaptive purification trigger in the eight-node chain. The no-purification, predictive-adaptive, and fixed-purification policies are nearly trigger independent, while the threshold-adaptive policy changes sharply as the trigger crosses the elementary-link fidelity scale. (b) Above-target delivery probability as a function of the target-fidelity threshold for the eight-node CUDA-Q/SeQUeNCe event-layer simulation. Increasing the target threshold narrows the useful operating region for all policies, illustrating the difficulty of satisfying high end-to-end fidelity in a longer repeater chain. (c) Target-fidelity sweep for the clean eight-node stationary configuration. The sweep shows the transition from a low-target regime, where weaker purification can still deliver above-target pairs, to a high-target regime where all policies lose useful delivery probability. Error bars denote 95% confidence half-widths over Monte Carlo trials.

Fig 9(b) demonstrates that purification collapses rapidly as the threshold is raised, showing that raw delivery alone is insufficient in the longer chain. This establishes that fixed purification remains useful over a wider target range, but its benefit eventually disappears as the target becomes too strict. The sweep identifies the intermediate target-fidelity regime in which adaptive purification is meaningful.

Fig. 9(c) provides the corresponding clean eight-node target-fidelity sweep. The same operating-regime structure is visible: low thresholds are accessible to multiple policies, intermediate thresholds separate the adaptive and fixed-purification strategies, and high thresholds drive all above-target probabilities toward zero. The nonmonotonic behavior of the predictive-adaptive curve reflects the discrete nature of mask selection and the fact that small changes in target fidelity can switch the selected purification pattern. This figure supports the main-text conclusion that useful entanglement delivery is not a monotonic function of purification effort; instead, it depends on balancing fidelity recovery against purification-induced failures and latency.

Fig. 10 decomposes the eight-node stationary policy comparison into fidelity, raw delivery, and latency components. Fig.10(a) shows that fixed purification gives the highest mean delivered fidelity, followed by the risk-aware predictive policy. However, fidelity alone does not determine useful entanglement delivery. Although fixed purification produces the largest delivered fidelity, it also introduces more purification attempts and therefore more opportunities for purification failure. The risk-aware policy achieves a slightly lower mean delivered fidelity than fixed purification, but remains above the target threshold while avoiding unnecessary purification on some links.

Fig. 10(b) shows the corresponding raw delivery tradeoff. No purification has the highest raw delivery probability because it avoids purification failures. However, as shown in the main text, those delivered pairs are not useful because their fidelities are below the target. Fixed purification has the lowest raw delivery probability because every link is purified, increasing the chance that a request fails before final delivery. The risk-aware predictive policy lies between these extremes: it sacrifices some raw delivery relative to no purification and mean-field prediction, but preserves substantially more fidelity-qualified delivery than those policies.

Fig. 10(c) shows that latency increases with purification effort. No purification has the lowest mean delivered latency, while fixed purification has the highest latency. The risk-aware predictive policy again occupies an intermediate operating point, with higher latency than threshold-adaptive and mean-field predictive purification but lower latency than fixed purification. Together, the three figures explain the main eight-node result: fixed purification maximizes delivered fidelity but pays a large delivery and latency cost, while risk-aware predictive purification better balances fidelity recovery, delivery probability, and resource cost. This is why the risk-aware policy achieves the largest above-target delivery probability in the main-text comparison.

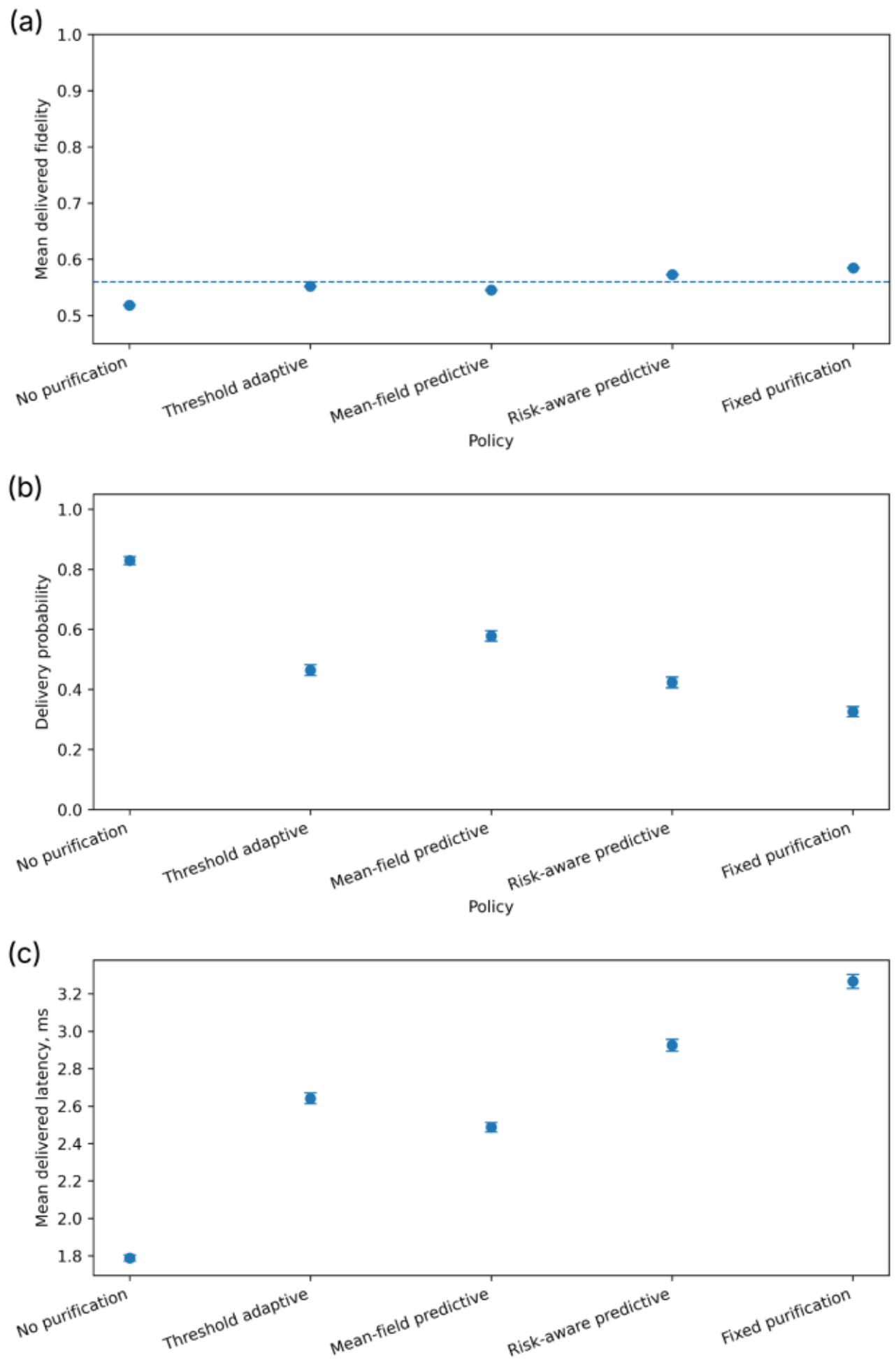

FIG. 10. (Colour online) Supporting eight-node stationary diagnostics for the risk-aware purification comparison. (a) Mean delivered fidelity for each purification policy. The dashed horizontal line marks the target-fidelity threshold. No purification, threshold-adaptive purification, and mean-field predictive purification remain at or below the target on average, while risk-aware predictive purification and fixed purification exceed the threshold. (b) Raw delivery probability for each policy. No purification gives the highest raw delivery probability, while fixed purification gives the lowest raw delivery probability because every elementary link is purified and therefore introduces additional failure modes. (c) Mean delivered latency for each policy. Latency increases with purification effort, with no purification producing the lowest latency and fixed purification producing the highest latency. Error bars denote 95% confidence half-widths over Monte Carlo trials or delivered-sample uncertainty, as appropriate.

*IDS-aware control*: Tables A3 and A4 present full-trace analysis of IDS-aware data for benign only and attack only traffics respectively.

Table A3. Summary of Full Trace: SSDP IDS-aware Quantum Network

| Scenario | Delivery | Above-target | Mean fidelity | Latency (ms) | Purifications |
|---|---|---|---|---|---|
| Clean | 0.6762 ± 0.0084 | 0.6283 ± 0.0086 | 0.5376 | 2.272 | 1.891 |
| Attack unaware | 0.6766 ± 0.0084 | 0.2908 ± 0.0081 | 0.5148 | 2.660 | 1.891 |
| IDS aware | 0.4700 ± 0.0089 | 0.4470 ± 0.0089 | 0.5423 | 3.174 | 3.524 |
| Oracle aware | 0.4663 ± 0.0089 | 0.4443 ± 0.0089 | 0.5418 | 3.116 | 3.481 |

Table A4. Attack Period Only: SSDP IDS-aware Quantum Network

| Scenario | Delivery | Above-target | Mean fidelity | Latency (ms) | Purifications |
|---|---|---|---|---|---|
| Clean | 0.6758 ± 0.0105 | 0.6274 ± 0.0109 | 0.5375 | 2.265 | 1.890 |
| Attack unaware | 0.6776 ± 0.0105 | 0.0983 ± 0.0067 | 0.5018 | 2.883 | 1.890 |
| IDS aware | 0.3620 ± 0.0108 | 0.3436 ± 0.0107 | 0.5449 | 4.086 | 4.387 |
| Oracle aware | 0.3459 ± 0.0107 | 0.3354 ± 0.0106 | 0.5463 | 4.081 | 4.400 |

Table A3 shows that IDS-aware improvement over attack-unaware is 0.1562 ± 0.0120 in above-target probability. IDS-aware and oracle-aware differ by only 0.0028 ± 0.0126, so they are statistically indistinguishable at the Monte Carlo resolution.

Table A4 shows that during the attack period, IDS-aware control improves above-target delivery over attack-unaware by 0.2453 ± 0.0126. The result is statistically meaningful at the Monte Carlo level. IDS-aware and oracle-aware are statistically close: the difference is 0.0082 ± 0.0151.

The tradeoff is explicit. IDS-aware reduces raw delivery from 0.6776 to 0.3620, increases latency from 2.883 ms to 4.086 ms, and increases mean purifications from 1.890 to 4.387. This cost is not a weakness; it is the mechanism by which useful above-target delivery is recovered.

Fig. 11 summarizes the complete quantum-network and IDS-aware control workflow used in this study. The left side of the pipeline defines the stationary quantum-repeater model. We consider both a three-link baseline and an eight-node linear chain, with each elementary link described by a generation probability, raw fidelity, attempt time, and memory coherence time. CUDA-Q is used at the primitive level to validate Bell-pair generation, entanglement purification, and entanglement swapping. The resulting primitive estimates are cached and passed to the event-layer simulator rather than repeatedly evaluating the same noisy quantum kernels during every Monte Carlo trial.

In Fig. 11, the central part of the pipeline represents the SeQUeNCe-style event-layer simulation. For each request, the simulator samples elementary link generation, records waiting times, applies memory decay, executes the selected purification mask, samples purification and swapping outcomes, and records whether the request is delivered above target, delivered below target, or fails during generation, purification, or swapping. This event-layer structure allows the policies to be evaluated using network-level performance metrics such as raw delivery probability, above-target delivery probability, delivered fidelity, latency, purification count, and failure-stage statistics.

The lower-left branch defines the stationary purification policies. No purification, threshold-adaptive purification, mean-field predictive purification, risk-aware predictive purification, and fixed purification are compared under identical Monte Carlo conditions. The risk-aware policy evaluates candidate masks using an inner Monte Carlo planner and selects a mask that balances above-target delivery, fidelity margin, purification cost, and latency cost.

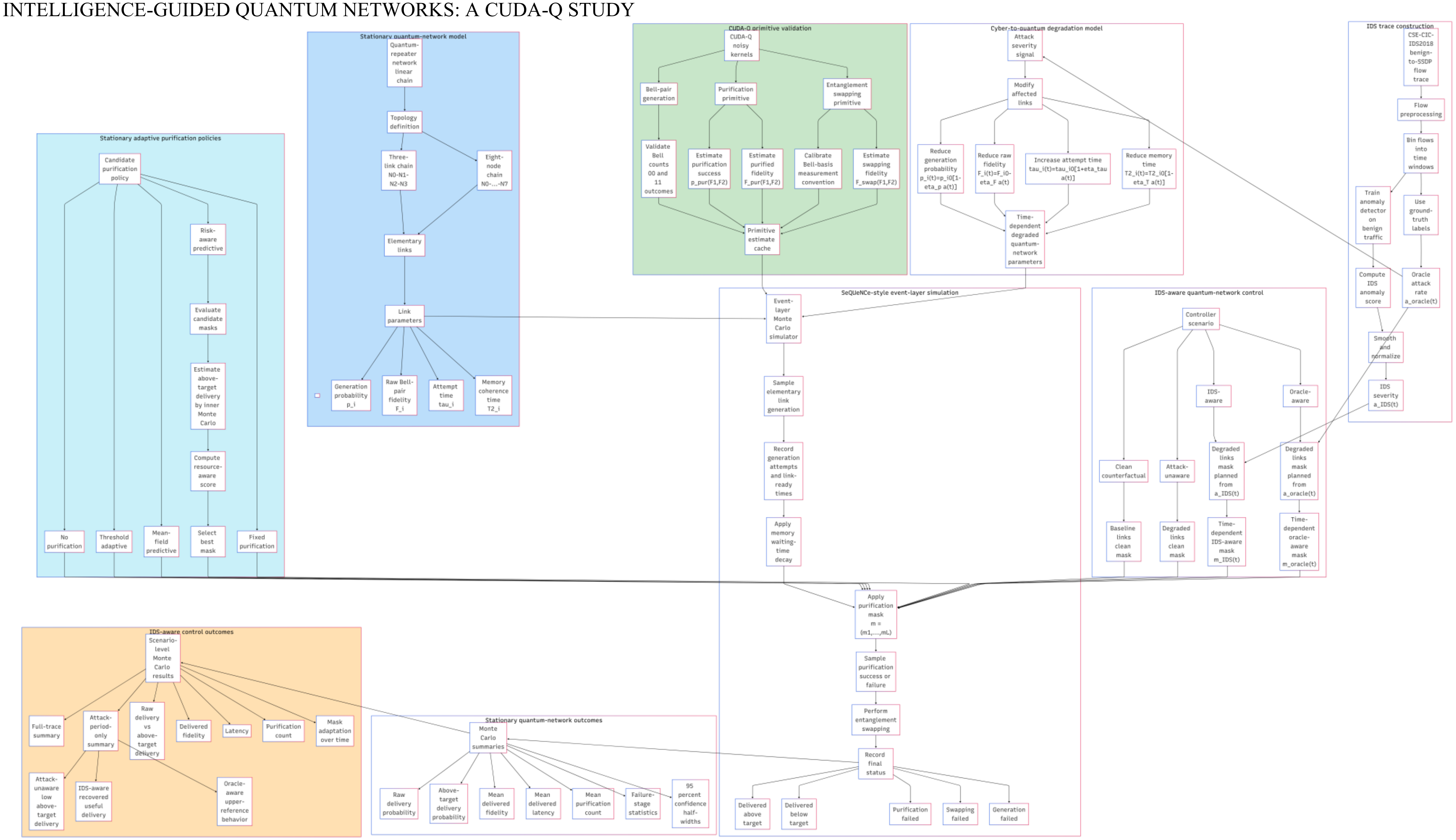

FIG. 11. (Colour online).Full quantum-network and IDS-aware control pipeline. The workflow begins with a stationary linear quantum-repeater model, including the three-link baseline and the eight-node chain. Each elementary link is parameterized by its generation probability ($p_i$), raw Bell-pair fidelity ($F_i$), attempt time ($\tau_i$), and memory coherence time ($T_{2,i}$). CUDA-Q noisy kernels validate the primitive Bell-pair generation, purification, and entanglement-swapping operations, and the resulting success-probability and fidelity estimates are stored in a primitive-estimate cache. These estimates are then supplied to a SeQUeNCe-style event-layer Monte Carlo simulator, which samples link generation, memory waiting-time decay, purification success or failure, entanglement swapping, and final delivery status. Stationary purification policies include no purification, threshold-adaptive purification, mean-field predictive purification, risk-aware predictive purification, and fixed purification. For IDS-aware control, a CSE-CIC-IDS2018 benign-to-SSDP flow trace is converted into an IDS severity signal and an oracle attack-rate reference. The attack severity drives a cyber-to-quantum degradation model that modifies affected link parameters, while the controller selects clean, attack-unaware, IDS-aware, or oracle-aware purification masks. The final outputs include raw delivery probability, above-target delivery probability, delivered fidelity, latency, purification count, mask adaptation, and failure-stage statistics. Produced with Mermaid [82].

The right side of the pipeline in Fig. 11 introduces cyber-aware control. A CSE-CIC-IDS2018 benign-to-SSDP trace is binned into time windows and used to construct two cyber-state signals: an IDS severity score obtained from anomaly detection and an oracle attack rate computed from ground-truth labels. The oracle signal defines the physical degradation applied to selected quantum links, while the IDS signal is used by the deployable controller to select time-dependent purification masks. This separation enables comparison among clean, attack-unaware, IDS-aware, and oracle-aware scenarios. The final IDS-aware outputs measure whether cyber-state information changes the purification action and improves fidelity-qualified entanglement delivery under SSDP-induced degradation.

## APPENDIX B: MATHEMATICAL FORMULATIONS

**Proposition 1: Existence of an empirical optimal mask**

**Statement:** For a finite repeater chain with $L$ elementary links, the empirical optimization problem

$$\hat{m} = \arg \max_{m \in \{0,1\}^L} J(m)$$

has at least one solution.

**Proof:** The set of all purification masks is

$$\mathcal{M} = \{0,1\}^L.$$

This set is finite, with

$$|\mathcal{M}| = 2^L.$$

The empirical score $J(m)$ assigns a real number to each mask $m \in \mathcal{M}$.

A real-valued function over a finite set always attains a maximum. Therefore, there exists at least one mask

$$\hat{m} \in \mathcal{M}$$

such that

$$J(\hat{m}) \geq J(m)$$

for every

$$m \in \mathcal{M}.$$

Therefore, the empirical optimization problem has at least one solution.

**Proposition 2: Empirical dominance over included baselines**

**Statement:** Suppose a baseline policy corresponds to a mask

$$m_0 \in \mathcal{M}.$$

If the risk-aware controller chooses

$$\hat{m} = \arg \max_{m \in \mathcal{M}} J(m),$$

then

$$J(\hat{m}) \geq J(m_0).$$

**Proof:** By definition, $\hat{m}$ maximizes $J(m)$ over all masks in $\mathcal{M}$. Therefore,

$$J(\hat{m}) \geq J(m)$$

for every mask

$$m \in \mathcal{M}.$$

Since the baseline mask $m_0$ is also an element of $\mathcal{M}$,

$$m_0 \in \mathcal{M},$$

we obtain

$$J(\hat{m}) \geq J(m_0).$$

Thus, the selected risk-aware mask is never worse than the baseline mask under the empirical planning objective.

*Note*: This does **not** guarantee that $\hat{m}$ will always outperform $m_0$ in a separate Monte Carlo evaluation. It only guarantees that $\hat{m}$ is at least as good as $m_0$ according to the empirical planning score $J$. Evaluation performance can differ because of finite sampling noise and stochastic event trajectories

**Theorem 1: Uniform concentration of above-target delivery estimates**

**Statement:** Let

$$U(m) = \Pr\left(D_m = 1 \text{ and } F_m \geq F_\star\right)$$

be the true above-target delivery probability for mask $m$.

Let

$$\hat{U}_n(m) = \frac{1}{n}\sum_{j=1}^{n} Y_j(m)$$

be the empirical estimate from $n$ independent trials.

Then, with probability at least $1-\delta$,

$$\sup_{m \in \mathcal{M}} \left|\hat{U}_n(m) - U(m)\right| \leq \sqrt{\frac{\log\left(2|\mathcal{M}|/\delta\right)}{2n}}.$$

**Proof:** For a fixed mask $m$, the variables

$$Y_1(m), Y_2(m), \ldots, Y_n(m)$$

are independent Bernoulli random variables with mean

$$U(m).$$

By Hoeffding's [84] inequality,

$$\Pr(|(m) - U(m)| \,|\, \epsilon) \leq 2\exp\left(-2n\epsilon^2\right).$$

This bound holds for a single fixed mask. We want a bound that holds for all masks simultaneously.

Using the union bound over all masks in $\mathcal{M}$,

$$\Pr\left(\exists m \in \mathcal{M} : \left|\hat{U}_n(m) - U(m)\right| > \epsilon\right) \le 2|\mathcal{M}|\exp\left(-2n\epsilon^2\right).$$

Set the right-hand side equal to $\delta$ :

$$2|\mathcal{M}|\exp\left(-2n\epsilon^2\right) = \delta.$$

Divide by $2|\mathcal{M}|$ :

$$\exp\left(-2n\epsilon^2\right) = \frac{\delta}{2|\mathcal{M}|}.$$

Take logarithms:

$$-2n\epsilon^2 = \log\left(\frac{\delta}{2|\mathcal{M}|}\right).$$

Equivalently,

$$2n\epsilon^2 = \log\left(\frac{2|\mathcal{M}|}{\delta}\right).$$

Therefore,

$$\epsilon^2 = \frac{\log(2|\mathcal{M}|/\delta)}{2n}.$$

Taking the square root gives

$$\epsilon = \sqrt{\frac{\log\left(2|\mathcal{M}|/\delta\right)}{2n}}.$$

Therefore, with probability at least $1-\delta$ ,

$$\sup_{m\in\mathcal{M}} |\hat{U}_n(m) - U(m)| \le \sqrt{\frac{\log\left(2|\mathcal{M}|/\delta\right)}{2n}}.$$

This theorem says that the empirical above-target delivery estimates become uniformly accurate over all candidate masks as the number of inner Monte Carlo planning trials increases. For the 8-node chain,

$$|\mathcal{M}| = 128,$$

so the number of candidate masks is small enough for exhaustive evaluation.

**Corollary 1: More planning trials improve mask-selection reliability**

**Statement:** For fixed mask set $\mathcal{M}$ and confidence level $1-\delta$ , the uniform error bound

$$\sqrt{\frac{\log\left(2|\mathcal{M}|/\delta\right)}{2n}}$$

decreases as the number of planning trials $n$ increases.

**Proof:** The numerator

$$\log\left(2|\mathcal{M}|/\delta\right)$$

does not depend on $n$ .

The denominator is $2n.$ Therefore, the full expression scales as

$$n^{-1/2}.$$

Hence, increasing $n$ decreases the uniform estimation error.

**Theorem 2: Near-optimality of the empirical mask**

**Statement:** Let $S(m)$ be the true resource-penalized score of mask $m$ , and let $\hat{S}(m)$ be its empirical estimate.

Assume that, with probability at least $1-\delta$ ,

$$\sup_{m\in\mathcal{M}} \left|\hat{S}(m) - S(m)\right| \le \epsilon.$$

Let

$$m^\star = \arg\max_{m\in\mathcal{M}} S(m)$$

be the true optimal mask, and let

$$\hat{m} = \arg\max_{m\in\mathcal{M}} \hat{S}(m)$$

be the empirical optimizer.

Then, with probability at least $1-\delta$ ,

$$S\left(m^\star\right) - S(\hat{m}) \le 2\epsilon.$$

**Proof:** Because $\hat{m}$ maximizes the empirical score,

$$\hat{S}(\hat{m}) \ge \hat{S}\left(m^\star\right).$$

By the uniform error assumption,

$$S\left(m^\star\right) \le \hat{S}\left(m^\star\right) + \epsilon.$$

Using the empirical optimality of $\hat{m}$ ,

$$\hat{S}\left(m^\star\right) \le \hat{S}(\hat{m}).$$

Therefore,

$$S\left(m^\star\right) \le \hat{S}(\hat{m}) + \epsilon.$$

Again by the uniform error assumption,

$$\hat{S}(\hat{m}) \le S(\hat{m}) + \epsilon.$$

Combining the inequalities,

$S\left(m^{\star}\right) \leq S\left(\hat{m}\right)+2\epsilon$. Rearranging gives

$$S\left(m^{\star}\right)-S\left(\hat{m}\right) \leq 2\epsilon.$$

Thus, the empirical optimizer is within $2\epsilon$ of the true optimizer under the score $S$.

If the empirical score estimates are accurate for every mask, then the selected empirical mask is nearly optimal for the true score. This supports exhaustive inner Monte Carlo planning over the finite mask set.

**Proposition 3: IDS-aware control can reduce raw delivery while improving useful delivery**

**Statement:** Let $m_{\text{unaware}}$ be the attack-unaware purification mask, and let $m_{\text{IDS}}$ be the IDS-aware mask. It is possible that

$$P_{\text{del}}\left(m_{\text{IDS}}\right) < P_{\text{del}}\left(m_{\text{unaware}}\right)$$

while

$$U\left(m_{\text{IDS}}\right) > U\left(m_{\text{unaware}}\right).$$

**Proof:** Raw delivery probability is

$$P_{\text{del}}\left(m\right) = \Pr\left(D_m = 1\right).$$

This counts all delivered pairs, regardless of fidelity.

Above-target delivery probability is

$$U\left(m\right) = \Pr\left(D_m = 1 \text{ and } F_m \geq F_{\star}\right).$$

This counts only delivered pairs whose final fidelity exceeds the target.

An attack-unaware controller may use a low-purification mask. That can give high raw delivery because fewer purification operations are attempted and therefore fewer purification failures occur. However, under attack-induced degradation, many delivered pairs may have

$$F_m < F_{\star}.$$

Therefore, raw delivery can be high while above-target delivery is low.

An IDS-aware controller may choose a more purification-heavy mask. This can reduce raw delivery because purification may fail and because additional operations increase latency. However, when the controller does deliver an end-to-end pair, the fidelity is more likely to exceed $F_{\star}$. Thus above-target delivery can increase even while raw delivery decreases. Therefore, it is possible for

$$P_{\text{del}}\left(m_{\text{IDS}}\right) < P_{\text{del}}\left(m_{\text{unaware}}\right)$$

and

$$U\left(m_{\text{IDS}}\right) > U\left(m_{\text{unaware}}\right)$$

to hold at the same time.

Example: During the SSDP attack period, the attack-unaware controller gives

$$P_{\text{del}} \approx 0.678,$$

The IDS-aware controller gives lower raw delivery,

$$P_{\text{del}} \approx 0.362,$$

Also, during the SSDP attack period,

$$U_{\text{unaware}} = 0.098 \pm 0.007,$$

while

$$U_{\text{IDS}} = 0.344 \pm 0.011.$$

The absolute improvement is

$$0.344 - 0.098 = 0.246.$$

The relative improvement is approximately

$$\frac{0.344}{0.098} \approx 3.5.$$

Therefore, during the SSDP attack period, IDS-aware control improves above-target entanglement delivery by about a factor of 3.5 relative to the attack-unaware controller. The *oracle-aware* result is

$$U_{\text{oracle}} = 0.335 \pm 0.011.$$

Thus, the IDS-aware result is close to the oracle-aware result:

$$U_{\text{IDS}} \approx U_{\text{oracle}}.$$

Thus the IDS-aware controller sacrifices raw throughput to recover fidelity-qualified useful entanglement. This supports the claim that the learned IDS anomaly score provides a useful cyber-state signal for adaptive quantum-network control.